\documentclass[apj]{emulateapj}
\slugcomment{{\sc Accepted to ApJS:} November 3, 2011}
\pdfoutput=1
\usepackage{graphicx,natbib,avant,color,psfrag,amsmath}
\usepackage{multirow, array,booktabs,mdwlist}

\shorttitle{IFS reduction, Spectral Extraction}
\shortauthors{Pueyo et.al}

\begin{document}

\title{Application of a damped Locally Optimized Combination of Images method to the spectral characterization of faint companions using an Integral Field Spectrograph}


\author{Laurent Pueyo  \altaffilmark{1}  \altaffilmark{2} \altaffilmark{*}, Justin R. Crepp \altaffilmark{3}, Gautam Vasisht\altaffilmark{4}, Douglas Brenner \altaffilmark{6}, Ben R. Oppenheimer \altaffilmark{6}, Neil Zimmerman \altaffilmark{6}, Sasha Hinkley \altaffilmark{3}, Ian Parry \altaffilmark{7}, Charles Beichman \altaffilmark{5}, Lynne Hillenbrand \altaffilmark{3}, Lewis C. Roberts Jr. \altaffilmark{4}, Richard Dekany \altaffilmark{3}, Mike Shao \altaffilmark{4}, Rick Burruss \altaffilmark{4}, Antonin Bouchez \altaffilmark{3}, Jenny Roberts \altaffilmark{3}, R\'{e}mi Soummer \altaffilmark{2}}

\altaffiltext{1}{NASA Sagan Fellow, Johns Hopkins University, Department of physics and astronomy,  366 Bloomberg Center 3400 N. Charles Street, Baltimore, MD 21218 USA}

\altaffiltext{2}{Space Telescope Science Institute, 3700 San Marin Drive, Baltimore, MD 21218, USA}
   
  \altaffiltext{3}{California Institute of Technology, 1200 E. California Blvd., Pasadena, CA 91125} 

 \altaffiltext{4}{Jet propulsion Laboratory, California Institute of technology, 4800 Oak Grove Drive, Pasadena, CA 91109 , USA} 
  
 \altaffiltext{5} {NASA Exoplanet Science Institute, 770 S. Wilson Avenue, Pasadena, CA 91225}
 
 \altaffiltext{6}{American Museum of Natural History, Central Park West at 79th Street, New York, NY 10024}
 
\altaffiltext{7}{University of Cambridge, Institute of Astronomy, Madingley Rd, Cambridge, CB3 0HA, UK}
  
\altaffiltext{*}{Corresponding author: lap@pha.jhu.edu}


\begin{abstract}
High-contrast imaging instruments are now being equipped with integral field spectrographs (IFS) to facilitate the detection and characterization of faint substellar companions. Algorithms currently envisioned to handle IFS data, such as the Locally Optimized Combination of Images (LOCI) algorithm, rely upon aggressive point-spread-function (PSF) subtraction, which is ideal for initially identifying companions but results in significantly biased photometry and spectroscopy due to unwanted mixing with residual starlight. This spectro-photometric issue is further complicated by the fact that algorithmic color response is a function of the companion's spectrum, making it difficult to calibrate the effects of the reduction without using iterations involving a series of injected synthetic companions. In this paper, we introduce a new PSF calibration method, which we call ``damped LOCI", that seeks to alleviate these concerns. By modifying the cost function that determines the weighting coefficients used to construct PSF reference images, and also forcing those coefficients to be positive, it is possible to extract companion spectra with a precision that is set by calibration of the instrument response and transmission of the atmosphere, and not by post-processing. We demonstrate the utility of this approach using on-sky data obtained with the Project 1640 IFS at Palomar. Damped-LOCI does not require any iterations on the underlying spectral type of the companion, nor does it rely upon priors involving the chromatic and statistical properties of speckles. It is a general technique that can readily be applied to other current and planned instruments that employ IFS's.
\end{abstract}

\maketitle 

\section{Introduction}

Current instruments dedicated to the imaging of exo-planets have a contrast floor that is set by the imperfections on the surfaces of the the optical train. In the case of a space observatory, as shown for instance by \cite{2005AJ....130.2778K,2008Sci...322.1345K}, when the optical surfaces corrugations are stable in time, the instrument response can be calibrated and subtracted. However, in the ground based regime, the PSF varies over timescales that range from minutes to hours e.g \cite{2007ApJ...654..633H,2006ApJ...637..541F}. Since these so called quasi-static speckles are the limiting factor of ground and spaced-based observations with regard to faint companion detectability, their properties have been studied extensively over the past few years. It was found that their theoretical properties (\cite{2003ApJ...596..702P,2004ApJ...612L..85A,2007ApJ...663L..49S,2007ApJ...669..642S}) match the statistics observed on-sky (\cite{2006ApJ...637..541F,2007ApJ...654..633H}).  

Several observing techniques and data reduction algorithms have been devised over the past few years in order to optimize the sensitivity of current imaging instruments to faint exo-planets. Observations are generally carried out by acquiring of a series of images in which the observer deliberately chooses to introduce some diversity in the imaging method. The purpose of this diversity is to force the image location of the residual speckles to vary from frame to frame, while the image of the sky is immobile. This is the case in the Angular Differential Imaging  (\cite{2006ApJ...641..556M,2007lyot.confE..44L}), Spectral Differential Imaging (\cite{1999PASP..111..587R,2004ApJ...615L..61M,2007ApJS..173..143B}), chromatic differential imaging (\cite{2002ApJ...578..543S}) and Coherent Imaging (\cite{2010PASP..122...71G}). Post-processing techniques involve a manipulation of this series of images whose purpose is to remove the artifacts due to speckles and leave intact the signal of a companion. The powerful combination of ADI data processed using a  Locally Optimized Combination of Images (LOCI) PSF subtraction algorithm allowed \citet{2008Sci...322.1348M} to obtain one of the first images of a planetary system. The LOCI PSF subtraction technique was devised by \citet{2007ApJ...660..770L} and is quickly becoming a standard in exo-planet searches (\citet{2008Sci...322.1348M,2009ApJ...707L.123T,2009ApJ...705L.204M,JustinSpeckle}). Other techniques currently discussed in the literature involve a-priori models of the PSF and the telescope \cite{2009JOSAA..26.1326M,Burke:10} or spectral deconvolution based on priors about the underlying spectrum of the companion \cite{2011arXiv1103.4766M}. On the contrary, LOCI only requires minimal priors about the behavior of the instrument and builds a composite reference PSF solely on the data obtained during observations.

In a previous paper we reported a recent implementation of LOCI that is applied to process data from an Integral Field Spectrograph (IFS), \citet{JustinSpeckle}. Our results were obtained using Project 1640. The full system  is composed of the Palomar Adaptive Optics (PALAO) followed by a stellar coronagraph (\cite{2005ApJ...618L.161S}), and the science science IFS. Further details about the instrument are found in \citet{2011PASP..123...74H}. The raw data from the infrared detector is calibrated and re-arranged in order to produce a data cube of images at $23$ different wavelengths across the J and H bands ($R \sim 40$). The spatial and spectral data extraction procedures are detailed in \citet{NeilPipeline}. The P1640 configuration prevents the field rotation that is necessary to  carry out ADI observations, but allows both chromatic and temporal diversity. Using an IFS behind a high contrast starlight suppression system is key to the two main science objectives of P1640: (a) detection of faint companions around nearby stars, (b) in-situ low resolution spectral characterization of any companions discovered.

The detection of faint point sources is facilitated by the chromatic diversity of the IFS and allows the observer to apply advanced static speckles calibration methods. Indeed, if all the errors are in an instrument pupil plane, diffractive optics predicts that the speckle pattern will stretch with wavelength, while the scale of the astronomical image is independent of wavelength. In an IFS, the series of measurements of the quasi-static speckle pattern is simultaneous. This alleviates the problem of corrugations evolving temporally between sequential observations. On the P1640 camera the speckles located at $1''$ are at the $5 \times 10^{-3}$ level and de-correlate in intensity by roughly $1$ percent per minute. This implies that purely sequential observations, separated by $5$ minutes in time, can at best reach a contrast of $2.5 \times 10^{-4}$ after subtraction. Using an IFS alleviates this problem as it provides reference PSFs which are contemporaneous realizations of the speckle pattern present in the science exposure. Each wavelength slice can be interpolated in order to produce rescaled data cubes in which the speckle pattern's scale remains constant. In this rescaled space the companion moves radially. If the variations of the speckle's brightness as a function of wavelength is know a-priori, then these rescaled images can be subtracted one to another to detect a faint companion. This approach was first discussed in \citet{2002ApJ...578..543S} and has been successfully applied to both simulated and on-sky IFS data \citet{2008A&A...489.1345V,2007MNRAS.378.1229T} . It is very powerful as it uses reference images that have all been acquired simultaneously, thus reducing the impact of the temporal evolution of the speckles to time scales shorter than the  exposure time. In \citet{JustinSpeckle} we presented an algorithm that combines the chromatic differential imaging in \citet{2002ApJ...578..543S} with the LOCI PSF subtraction approach. While LOCI was first presented in \citet{2007ApJ...660..770L} as a method to reduced ADI images, P1640 does not take advantage of  field rotation. Instead, the baseline P1640 reduction presented in \citet{JustinSpeckle} relies on the IFS' radial chromatic stretch of the PSF  and on temporal diversity. It integrates them both in a LOCI-type reduction scheme.We showed that this quasi-static speckles calibration method provided a gain greater than an order-of-magnitude in the contrast between raw and processed images, and an ultimate detection level $\simeq 1\times 10^{-5}$ at $1''$ in a $20$ minutes H-band exposure on a three magnitude star. Further improvements in contrast, while maintaining reasonable exposure times, require coherent wavefront calibrations, such as the ones presented in \citet{2006ApJ...638..488B,2010PASP..122...71G,Giveon:07,Pueyo:09}, for the case of space-based observations and \citet{wallace:74400S,2007JOSAA..24.2334S} for the case of ground-based telescopes.

A second advantage of the P1640 IFS is that, once a companion is detected by means of the aforementioned chromatic diversity, its spectral information is an immediate by-product. Over the past ten years, IFS have been used to characterize companions with favorable contrast ratio (e.g. \citet{2007ApJ...656..505M,2010A&A...517A..76P}). When the companions is fainter than the speckle floor, then reductions such as the ones presented in \citet{JustinSpeckle} ought to be carried out to first detect the companion. However, blind applications of  LOCI post-processing tailored for detections to spectral data-cubes present the major caveat of altering the spectro-photometry and possibly introduce artificial features in the low resolution of the detected object. Recent results obtained using the OSIRIS IFS at the Keck observatory, show a depletion of sixty percent of the companion flux that is constant over the spectral window (\cite{2010arXiv1008.4582B}). In this example the depletion is gray and as such has little impact on the scientific conclusions of the observations. However P1640 seeks for faint objects at close angular separations whose signal is buried within the bright speckle field, thus requiring more aggressive PSF subtraction. Moreover P1640 entirely relies on chromatic diversity, which can potentially lead to a poor conditioning of the inverse problem associated with LOCI (see below). In practice, when testing our reduction on synthetic objects injected within the control radius of the AO system, we observe that the post-LOCI retrieved spectral information is considerably biased with a non-gray algorithmic response. This biasing of the spectral information is entirely due to the PSF subtraction algorithm, and can be an order-of-magnitude larger than the next most important uncertainty, due to the telluric calibration described in \citet{NeilPipeline}. It stems from two independent phenomena: a bias caused by partial fitting of the companion flux with the starlight in the reference frames, and a spectral cross-talk between actual companion flux between neighboring channels. The combination of these errors is very often the dominant source of uncertainties in spectroscopic measurements carried out using post-LOCI subtracted IFS data-cubes.

Herein we introduce a new method that lowers the error in the measured Spectral Energy Distribution due to the reduction algorithm close to the level of uncertainty in the telluric calibration. We first introduce a modified version of the LOCI algorithm that alleviates the problem of flux depletion of faint companions  that is common to most locally optimized PSF subtraction algorithms. The ``damped LOCI'' algorithm introduced in this communication seeks to maximize the residual flux of a putative companion while minimizing the noise due to the quasi-static speckles, whereas standard LOCI implementations only focus on the latter objective. This is achieved by re-formulating the least-squares problem associated with finding a locally optimal composite reference PSF: we augment the cost function associated with the determination of the composite PSF coefficients with a penalty term that scales with the flux of a potential companion. ``Damped LOCI'' thus solves the following optimization problem ``{\it minimize the least-squares fit to the local speckle field while maximizing the least-squares fit in the area where a known companion is present}".  

The formalism and the notations underlying LOCI and "damped LOCI" are details in appendix A. \S  2 
focuses on the specifics and the implementation of each algorithm. Both approaches yield comparable performances in terms of band-averaged  faint companions' SNR. Appendix B discusses the second order impact of each method on the detectability problem, and in particular  the advantages of "damped LOCI" when the inverse problem associated with the PSF reconstruction is ill-conditionned. In \S 3  we discuss  the core matter of this paper, namely the measurement of  the spectrum of faint companions. When companions have already been detected we show that a series of binary PSF weightings, or masking, can be used in conjunction with classical LOCI methods to retrieve the spectrum of relatively bright objects. We illustrate however that the performance of such methods degrades rapidly with increasing contrast, and show that using masking in conjunction with ``damped LOCI'' decouples spectral precision and contrast. We show that  this method yields  in all cases better absolute and relative spectro-photmetric accuracy when compared with classical LOCI reductions.  In particular, for objects as bright as the residual speckles, we obtain estimates of the Spectral Energy Distribution of the order of  the telluric calibration uncertainties of the spectrograph . For faint objects ``damped LOCI'' yields SED uncertainties consistent with higher SNR (less ``photon starved'') state of the art band-averaged photometric estimates (\citet{2008Sci...322.1348M,2009ApJ...707L.123T,2009ApJ...705L.204M}.  As the architecture of future exo-planet imagers and P1640 are similar (\cite{2007arXiv0704.1454G,2006SPIE.6269E..24D}), the results presented here are broadly relevant to the future exo-planet imaging endeavors.

\section{Standard LOCI and damped LOCI implementations}

\subsection{Standard LOCI}
The Locally Optimized Combination of Images consists of reconstructing a reference PSF from IFS data. It is based on an inverse problem which seeks minimize the residuals between a given astronomical image, the target image to be reduced, and a calibration image. The algorithm, as introduced by \citet{2007ApJ...660..770L}, seeks to find an optimal combination of images, within an ensemble of reference frames, that locally minimizes the least-squares residual between the target frame and the composite reference, in a large optimization zone of the image, $\mathcal{O}$ zone.  This weighted reference is then subtracted from the reduced image for only a small subset of pixels, the subtraction zone ,$\mathcal{S}$ zone. \citet{JustinSpeckle} presented how such an approach can be utilized to reduced IFS data. Appendix A details the specifics of the frame selection and the notations underlying the present manuscript. In this section we focus on the kernel of the LOCI algorithm, namely the inverse problem associated with the determination of the optimal reference weights.

The search for the coefficients of the composite reference is based on a least square cost function integrated over the zone $\mathcal{O}$.
\begin{eqnarray}
& & min_{ \{ c_k \} } \left \{ \; \int_{\mathcal{O}} dx dy \; w(x,y) \left[T(x,y) - \sum_{k}^{N_{\mathcal{R} }} c_k  R_k(x,y)\right]^2 \right \} \nonumber \\
&\Leftrightarrow &min_{ \{ c_k \} } \left \{ ||T - \sum_{k}^{^{N_{\mathcal{R} }} }  c_k  R_k ||_{w,\mathcal{O}} \right \}
\label{Eq:IntCost}
\end{eqnarray}
where $T$  stands for the target image, $\{ R_{k} \}$ is the ensemble of reference images, $\{ c_{k} \}$ the ensemble of LOCI coefficients, $w(x,y)$ is a weighting function that can be used to mask certain pixels, $N_{\mathcal{R}}$ the number of reference frames in $\mathcal{R}$, and $ ||. ||_{w,\mathcal{O}}$ denotes the $L^{2}$ norm, weighted by $w$, over the optimization zone. In matrix notation we re-write Eq.~\ref{Eq:IntCost} as:
\begin{equation}
||T - \sum_{k}^{N_{\mathcal{R} }}  c_k  R_k ||_{w,\mathcal{O}} = c^{T} \; \mathbf{M}^{(R,R)}_{\mathcal{O}} \; c - 2 \mathbf{V}^{(R,T)}_{\mathcal{O}} \; c+ \mathbf{S}^{(T,T)}_{\mathcal{O}}
\label{Eq::CostSquare}
\end{equation}
where $c$ is a  $N_{\mathcal{R}}$ dimensional vector, $\mathbf{M}^{(R,R)}_{\mathcal{O}}$ is an $N_{\mathcal{R}} \times N_{\mathcal{R}}$ matrix, $\mathbf{V}^{(R,T)}_{\mathcal{O}}$ is a $N$ dimensional vector, and $\mathbf{S}^{(T,T)}_{\mathcal{O}}$ a scalar. They are defined by:
\begin{eqnarray}
c &=& [c_1, c_2, ..., c_k, ... c_{N_{\mathcal{R}}}]^{T}\\
\mathbf{M}^{(R,R)}_{\mathcal{O}} [p,q] &=& \int_{\mathcal{O}} dx dy \;w(x,y) R_{p}(x,y) \;R_{q}(x,y)\\
\mathbf{V}^{(R,T)}_{\mathcal{O}}[p] &=& \int_{\mathcal{O}}dx dy \;w(x,y)  R_{p}(x,y)\; T(x,y) \\
 \mathbf{S}^{(T,T)}_{\mathcal{O}} &=& \int_{\mathcal{O}}dx dy \; w(x,y) T(x,y) \; T(x,y) 
\end{eqnarray}
Using these notations, the quadratic residual minimization problem associated with standard LOCI implementations can be solved by taking the derivative of of Eq.~\ref{Eq::CostSquare}with respect to $c$,
\begin{eqnarray}
& &min_{ \{ c_k \} } \left \{||T - \sum_{k} c_k  R_k ||_{w,\mathcal{O}} \right \} \nonumber \\
&\Leftrightarrow&  \mathbf{M}^{(R,R)}_{\mathcal{O}} \; c - \mathbf{V}^{(R,T)}_{\mathcal{O}} = 0
\label{Eq::InverseProb}
\end{eqnarray}
If  the ensemble of references, $\{R_{k} \in \mathcal{R} \}$, forms a complete basis set of the $\mathcal{O}$-zone, then the matrix $\mathbf{M}^{(R,R)}_{\mathcal{O}}$ is invertible. If the speckle diversity is only temporal, and if each frame corresponds to an independent realization of the speckle pattern, then the inverse problem in Eq.~\ref{Eq::InverseProb} is well conditioned when the number of PSF cores in the optimization zone matches the number of reference images, $N_{A} = N_{\mathcal{R}}$. In this case the optimal weights are defined as
\begin{equation}
\boxed{ c_{LOCI} = (\mathbf{M}^{(R,R)}_{\mathcal{O}})^{-1} \mathbf{V}^{(R,T)}_{\mathcal{O}}}
\end{equation}
However when using IFS data, the color slices within each exposure correspond to the same realization of the speckle field, albeit seen at different wavelengths. As a consequence they are highly correlated and this results in poorer conditioning of $\mathbf{M}^{(R,R)}_{\mathcal{O}}$. This conditioning is a measure of how well  $N_A$ matches the number of linearly independent references frames in the zone $\mathcal{O}$. 

Carrying out a selection of the most relevant slices in $\mathcal{R}$, so that the number of independent references matches $N_{A}$, the number of PSF cores, ought to be done locally  and iteratively (\cite{2010SPIE.7736E..52M}). It represents a considerable computational overhead and in practice, LOCI reductions are carried out with $N_A$ constant over large portions of the image. Then, in the zones for which the problem is ill-posed, generic pseudo-inverse routines or eigenvalues truncations provide a robust alternative, at the cost of losing track of the physics in the process. Indeed, such inversion methods lead to the subtraction of the least correlated modes in the reference images. If the flux of the companion in the target frame can be described using some of these independent reference modes, then some of its flux is subtracted by the composite reference. The photometric information is then biased. Choosing an optimization zone much larger than the subtraction zone, a larger $N_{A}$, mitigates this issue. It increases the contribution of pixels that do not see any companion flux and reduces the likelihood of fitting modes associated with the companion with one of the reference modes. However this occurs at expense of loosing some of the local characteristics of the suppression. Such a procedure can be efficient outside of the control radius of the AO system, where the speckle halo is relatively uniform (\citet{2010arXiv1008.4582B}). However when the speckles are bright and structured, optimizing the LOCI parameters in order to obtain the best spectral retrieval quickly becomes an intractable problem. In other words, when the minimization is ill-posed,  pseudo-inverse routines aim at minimizing the speckle noise in the reduced image without considerations of the flux of a potential companion. Whereas such an approach is adequate for faint companion detection, it can be problematic when quantifying the spectro-photometry of the discovered point sources. We now explore a modification of the LOCI that resolves these issues.

\subsection{Flux preserving locally optimized speckle suppression: damped LOCI}
``Damped LOCI'' relies on the same zonal approach as above. The modification we introduce resides in posing differently the inverse problem that defines the modal coefficients of the composite reference. Instead of solely minimizing the residual least-squares fit between target frame and composite reference in the optimization zone, we constrain the possible combinations of coefficients over which the solution may be sought. We first impose positivity, $c_{k} \geqslant 0 $ for all $k$, in order to avoid the pathological case where two adjacent coefficients are large in absolute value but of different signs. Such a situation would lead to high spectral cross-talk from the companion flux at neighboring wavelengths. While this condition might appear too stringent, we show below that ensuring the positivity of all the reference coefficients allows us to introduce concisely in the minimization problem a penalty term that scales with opposite of the flux a companion. Because we are interested in devising a reduction method that does not modify the spectro-photometry of a detected companion, the quality of the fit in the zone $\mathcal{S}$ is of critical importance. Ideally, even if the flux from a companion is present in the zone $\mathcal{S}$, we want the algorithm to only fit the stellar contribution in that zone, leaving the companion contribution unperturbed. {\it In essence ``damped LOCI'' seeks composite references that are the result of a competition between minimizing the fit residual in the zone $\mathcal{O}$ and maximizing it in the zone $\mathcal{S}$, under the constraint that all the composite coefficients are positive}. Since the zone $\mathcal{O}$  is much larger than $\mathcal{S}$, the minimization of its residual fit should prevail, and as a consequence the speckles will be subtracted in a way that preserves more of the companion flux than in classical LOCI implementations.\\

Formally, the squared difference between $T$ and the composite reference integrated over $\mathcal{S}$ can be written as:
\begin{equation}
I_{S}^{(2)} = ||T - c R||_{w,\mathcal{S}} = c^{T} \; \mathbf{M}^{(R,R)}_{\mathcal{S}} \; c - 2 \mathbf{V}^{(R,T)}_{\mathcal{S}}\; c + \mathbf{S}^{(T,T)}_{\mathcal{S}}
\label{Eq::SquaredIntS}
\end{equation}
where we have replaced  the $\mathcal{O}$ subscript by the $\mathcal{S}$ subscript for the matrix and vectors. When there is flux from a  companion in the $\mathcal{S}$ zone, we want this quantity to remain large while the residual in the zone $\mathcal{O}$ is minimized. We can thus use an augmented cost function to write the optimization problem as follows:  
\begin{eqnarray}
& & min_{\{c_k,\Lambda \}} \left \{ J_{0}(c_k,\Lambda) \right \} \Leftrightarrow \label{Eq::CostLag} \\
& & min_{\{c_k,\Lambda \}} \left \{ c^{T} \; \mathbf{M}^{(R,R)}_{\mathcal{O}} \; c - 2 \mathbf{V}^{(R,T)}_{\mathcal{O}} \; c+ \mathbf{S}^{(T,T)}_{\mathcal{O}} \right.  \\
& & \left. - \Lambda  (c^{T} \; \mathbf{M}^{(R,R)}_{\mathcal{S}} \; c - 2 \mathbf{V}^{(R,T)}_{\mathcal{S}} \; c+ \mathbf{S}^{(T,T)}_{\mathcal{S}}) \right \} \nonumber
\end{eqnarray}
where $\Lambda$ is multiplicative constant that allows us to include the flux preservation constraint in the LOCI minimization. The term $- \Lambda  (c^{T} \; \mathbf{M}^{(R,R)}_{\mathcal{S}} \; c - 2 \mathbf{V}^{(R,T)}_{\mathcal{S}} \; c+ \mathbf{S}^{(T,T)}_{\mathcal{S}})$ is called the flux preservation penalty term and $- \Lambda$ is the Lagrange multiplier associated with this term. If one seeks to maximize the flux of a potential companion in the $\mathcal{S}$ zone by minimizing $J_{0}(c_k,\Lambda) $,  then $\Lambda$ ought to be positive. As a consequence, this cost function is not positive definite, making the practical implementation of this optimization difficult. Forcing positivity for the ${c_k}$ coefficients leads us to ensure the positivity of the second term in Eq.~\ref{Eq::CostLag} and thus facilitates the search for the optimal coefficients. \\

To do so, we project the LOCI processed image, $I^{P}$, over all the modes present in the collection of references. 
\begin{eqnarray}
 I^{P} &=& T - c R \label{Eq::BasicDiff} \\
\sum_{k =1}^{N_{\mathcal{R}}}  c_{k} \int_{\mathcal{S}} w R_{k} I^{P}  = \mathbf{V}^{(R,I^{P})}_{\mathcal{S}} \; c &=&\mathbf{V}^{(R,T)}_{\mathcal{O}} \; c- c^{T} \; \mathbf{M}^{(R,R)}_{\mathcal{S}} \; c
 \label{Eq::ProcessProjected}
\end{eqnarray}
where we have taken the inner product over the $\mathcal{S}$- zone of the composite reference with both sides of Eq.~\ref{Eq::BasicDiff}. Replacing the quadratic term  in Eq.~\ref{Eq::SquaredIntS} by the quadratic term in Eq.~\ref{Eq::ProcessProjected} yields
\begin{equation}
I_{S}^{(2)} = \mathbf{S}^{(T,T)}_{\mathcal{S}} - \mathbf{V}^{(R,I^{P})}_{\mathcal{S}} \; c - \mathbf{V}^{(R,T)}_{\mathcal{S}} \; c. 
\end{equation}
Since the condition $c_{k} \geqslant 0$ for all $k$ is being enforced and the actual images only include positive counts, then  $\mathbf{S}^{(T,T)}_{\mathcal{S}} > 0 $ and $\mathbf{V}^{(R,T)}_{\mathcal{S}} \; c >0$. In order to devise an augmented cost function that consists of the sum of positive scalar quantities, we need to show that maximizing $I_{S}^{(2)}$ is equivalent to minimizing of $\mathbf{V}^{(R,T)}_{\mathcal{S}} \; c$. When there is no companion in the $\mathcal{S}$  zone then $|\mathbf{V}^{(R,I^{P})}_{\mathcal{S}} \; c | \ll  |\mathbf{V}^{(R,T)}_{\mathcal{S}} \; c|$, since the flux in the processed image is smaller than the flux in the target image.  Even if  $\mathbf{V}^{(R,I^{P})}_{\mathcal{S}} \; c $ happens to be negative, which means that the subtraction worked too well, the integrated square flux in the subtraction zone is dominated by the behavior of  $\mathbf{V}^{(R,T)}_{\mathcal{S}} \; c$. When there is a companion then $\mathbf{V}^{(R,I^{P})}_{\mathcal{S}} \; c > 0$ and has the same sign as $\mathbf{V}^{(R,T)}_{\mathcal{S}} \; c$. Thus the integrated squared residual in the subtraction zone behaves as the opposite of the positive scalar $\mathbf{V}^{(R,T)}_{\mathcal{S}} \; c $. As a consequence we modify the augmented cost function shown in Eq.~\ref{Eq::CostLag}, in order to avoid the use of a negative Lagrange multiplier associated with the full quadratic form of the least-square fit in the $\mathcal{S}$ zone. Instead we choose to use $\mathbf{V}^{(R,T)}_{\mathcal{S}}$ in conjunction with a positive Lagrange multiplier as a penalty term.

Since  $\mathbf{V}^{(R,T)}_{\mathcal{S}}$  scales as the opposite of the residual fit in the $\mathcal{S}$ zone, this modification re-formulates the problem of minimizing Eq.~\ref{Eq::CostLag} in a tractable way. It allows us to write the following positive definite augmented cost function:
\begin{equation}
J_{1}(c,\Lambda) =  c^{T} \; \mathbf{M}^{(R,R)}_{\mathcal{O}} \; c - 2 \mathbf{V}^{(R,T)}_{\mathcal{O}} \; c+ \mathbf{S}^{(R,T)}_{\mathcal{O}} + \Lambda \mathbf{V}^{(R,T)}_{\mathcal{S}} \; c 
\end{equation}
where $\Lambda$ is the Lagrange multiplier associated with the quality of the least-squares fit in the $\mathcal{S}$ zone. By augmenting the cost function using a quantity that scales with the flux suppression of a potential companion, we introduce a damping term in the speckle suppression algorithm. This term forces the overall speckle reduction to compete with flux preservation in the $\mathcal{S}$ zone. Having chosen a positive definite cost function allows us to use standard quadratic programming routines to solve the following inverse problem:\\

\textbf{Find $c^{DL}$ and $\Lambda^{DL}$ that minimize $J(c^{DL},\Lambda^{DL}) $ under the constraint that $c^{ML}_k \geqslant 0$ for all $k$}\\

where the subscript $DL$ stands for ``damped LOCI'', d-LOCI hereafter. 

In practice we implement this optimization problem by nesting a linear least-squares solver within a one dimensional non-linear search,

\begin{equation}
\boxed{\Lambda^{DL}  =  min_{\Lambda} J(c(\Lambda),\Lambda)}
\end{equation}

with:
 
\begin{equation}
\boxed{c(\Lambda)  =  (\mathbf{M}^{(R,R)}_{\mathcal{O}})^{\dag} (\mathbf{V}^{(R,T)}_{\mathcal{O}} - \frac{\Lambda}{2} \mathbf{V}^{(R,T)}_{\mathcal{S}})}
\end{equation}

where the superscript $^{\dag}$ stands for the linear least square minimization under the constraint that all coefficients are positive. This algorithm will find the optimal weight between speckle and signal suppression and thus should behave better than regular LOCI with respect to the photometry and spectral features of a potential companion. Note that the damping in d-LOCI occurs in two separate instances: the positivity constraint on the coefficients and the conservation of the companion flux in the $\mathcal{S}$ zone.  We used a set of synthetic companions to assess the relative performances of both LOCI and d-LOCI with respect to detectability. Our results are presented in Appendix B, and illustrate that both methods  yield comparable SNR. In the next section we delve into the problem of  estimating of the spectro-photometry of  companions detected in IFS data using LOCI type reductions.
\section{Spectral extraction with d-LOCI}
\subsection{Spectral bias and contamination}
If we write the discrete spectrum of a companion, sampled at the spectral resolution of the IFS, as $\{I_{p}^{C}\}_{p = 1 \; ... \; N_{\lambda}}$, then the flux of a potential companion in  a LOCI subtracted image at the wavelength $\lambda_{p_0}$ can be written as
\begin{eqnarray}
I_{p_0}^{C,LOCI} &=&  I_{p_0}^{C}- I_{p_0}^{Bias}+ I_{Star}- I_{Cross}\\
I_{p_0}^{C,LOCI} &=&  I_{p_0}^{C}- I_{p_0}^{Bias}+ I_{Star}- \sum_{p \in \mathcal{R}} c_p \gamma_{p_0,p}  I_{p}^{C}.
\label{eq::problemspect}
\end{eqnarray} 
The three biases in Eq.~\ref{eq::problemspect} are described as follows:
\begin{itemize}
\item $ I_{p_0}^{Bias}$ corresponds to the amount of companion flux that is directly fitted by the PSF subtraction algorithm.  It is the portion of the companion light that has been considered as starlight by the LOCI algorithm and thus subtracted. This bias can occur in all implementations of LOCI, but is larger when using chromatic differential imaging, compared to ADI, because of the potential poor conditioning of the correlation matrix $\mathbf{M}^{(R,R)}_{\mathcal{O}}$. The relative size of the subtraction and optimization zones, which directly scales with $N_{A}$, impacts $ I_{p_0}^{Bias}$. This number ought to be carefully calibrated using synthetic point sources when measuring the photometry within any LOCI processed image. 
\item The bias $I_{Star}$ corresponds to the residual starlight after LOCI in the pixels of interest. If we assume that this residual background is constant around the companion PSF, it can be estimated and subtracted. As a consequence we neglect it here.
\item $I_{Cross}$ corresponds to the contamination of the measured flux at a given wavelength by the companion flux present in the composite reference at adjacent wavelengths. Indeed the measurement at $\lambda_{p_0}$ is contaminated by the spectral content of the companion at nearby wavelengths $ I_{p}^{C}$ multiplied by the corresponding LOCI coefficient, $c_p$, and a quantity, $\gamma_{p_0,p}$, that measures the spatial extend of the companion PSF between wavelength channels. $\mathcal{R}$ is chosen such that for all $p$ and $q$, $\gamma_{p,q} \ll 1$. However in the absence of damping the coefficients $|c_{p}|$ can become large enough to make the spectral contamination non negligible,$|c_{p}\gamma_{p,q} |\simeq 1$ . This spectral cross talk is specific to the use of LOCI with IFS images. It can be somewhat mitigated by optimizing the value of $N_{\delta}$ using synthetic companions. However such an optimization relies on knowing a priori the spectrum of the companion to characterize, and thus can only be carried out when this companion is bright enough to obtain a rough estimate of the spectrum in the absence of any  speckle calibration.
\end{itemize}
The final error on the retrieved spectrum is the sum of the algorithmic error, $I_{p_0}^{C} - I_{p_0}^{C,LOCI}$, and the spectroscopic calibration uncertainties. Calibrating the P1640 spectrograph involves a careful measurement of the instrumental spectral response function, whose determination procedure is detailed in \citet{NeilPipeline,2010ApJ...709..733Z,2010ApJ...712..421H}. The uncertainty on this response is not expected to be larger than $5$ percents. As we will see below, the quantity $ I_{p_0}^{C} - I_{p_0}^{C,LOCI}$ when using LOCI reduced data is much larger than this expected telluric calibration uncertainty and is thus very often the dominant source of errors. As a consequence we carried out our spectral fidelity computations in ``camera space'', using wavelength series that have not been normalized by the instrumental spectral response function. By doing so we chose to leave aside discussions about telluric calibration and only focus on issues related to the effects of the PSF subtraction algorithms.
\subsection{PSF binary weighting}
\begin{figure}
\begin{center}
\includegraphics[width=5in]{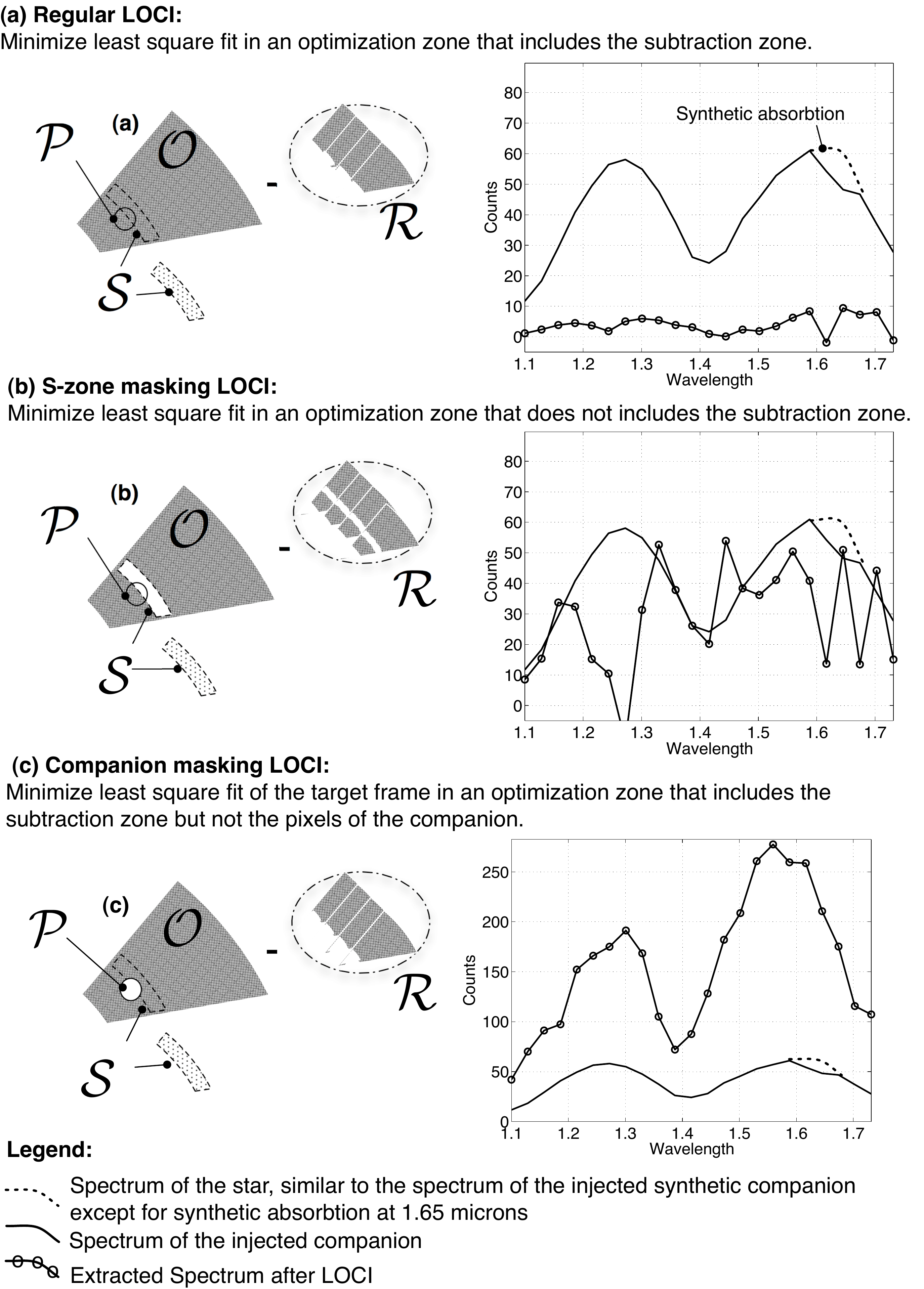}
\caption[Illustration of PSF masking to preserve the flux of the companion]{Illustration of three PSF binary weighting approaches, with their respective performances in the case of a companion slightly fainter than the P1640 speckles, $\Delta m = 6$. Note that the scale of the bottom panel has been compresses to accommodate for the large bias. None of these solution provide robust spectro-photometric accuracy: they all lead to combination of $I_{bias}$ and $I_{cross}$. $\mathcal{O}$ is the optimizations zone, used to find the best least-squares fit of the speckles realization. $\mathcal{S}$ is the subtraction zone. $\mathcal{P}$ is the position of the planet. $\mathcal{R}$ is the ensemble of reference images.}
\label{Fig::ZonalProcessExplainMasked1}
\end{center}
\end{figure}

In this section we consider the case of a faint companion which has been detected and needs to be characterized. We call $\mathcal{P}$ the image plane  location of the detected off-axis source, and assume the geometry described on the left hand side of Fig.~\ref{Fig::ZonalProcessExplainMasked1} and Fig.~\ref{Fig::ZonalProcessExplainMasked2}, where $\mathcal{P}$ and $\mathcal{S}$ have the same area but do not fully overlap. While the case of using as subtraction zone that is smaller than $\mathcal{P}$ is of particular interest for the detection problem, this configuration leads to large signal depletions and is thus not discussed in the present communication. The top panel of Fig.~\ref{Fig::ZonalProcessExplainMasked1} illustrates a test case carried out using aperture photometry on an IFS data-set reduced using the LOCI algorithm described in  \citet{2007ApJ...660..770L}. We introduced a fake companion located at  $1''$ in $4$ Alcor data-cubes. This corresponds to about $10$ minutes of exposure, each exposure being composed of $N_{\lambda} = 23$ frames (around $80$ reference). We ran our reductions with constant LOCI parameters $dr=2$, $N_{\delta} = 1$, $g = 3$ and $N_{A} =100$. This choice of $N_{A}$ leads to solving inverse problems which are slightly over-constrained and thus favorable to the regular LOCI in terms of detectability. In this figure the spectrum of the fake companion is similar to the one of the host star, except at $1.65 \mu m$, where a synthetic a $10 \%$  flux depletion was inserted, and its brightness was chosen such that its integrated delta magnitude over the J and H band was $\Delta m_{J+H} = 6.5$. The retrieved spectrum is very different from the injected one. Because the observed discrepancy cannot be described as a ``gray gain'', it is a combination of $I_{bias}$ and $I_{cross}$, it cannot be calibrated using a set of synthetic companions.   The $10 \%$ flux depletion at $1.65 \mu m$ allows us to illustrate the impact of $I_{cross}$. This weak absorption is  strongly amplified via wavelength cross-talk in the reduction method, as shown in the top panel of  Fig.~\ref{Fig::ZonalProcessExplainMasked1}. This aspect is very difficult to calibrate using synthetic sources since this depletion depends on the depth of the putative absorption, which is unknown a-priori. We first sought to address these issues by exploring several binary weighting strategies of the LOCI cost function:
\begin{itemize}
\item {\it Subtraction zone masking} consists on reducing $I_{bias}$ by not including the pixels in the $\mathcal{S}$ zone in the least square fit. It excludes the companion flux from the inverse problem and thus should preserve better the flux of potential companions. In practice it is implemented using a weighting function $w(x,y)$ that is $1$ in the optimization zone and $0 $ in the subtraction zone; and corresponds to solving the following optimization problem:
\begin{equation}
\mbox{Find } c_{k} \mbox{ st } min \left \{ ||T - \sum_{k}^{^{N_{\mathcal{R} }} }  c_k  R_k ||_{\mathcal{O} - \mathcal{S}} \right  \}
\end{equation}
The second panel of Fig.~\ref{Fig::ZonalProcessExplainMasked1} illustrates the efficiency of this approach in our test case. $I_{bias}$ is indeed reduced, however it is achieved at the cost of an increase in $I_{cross}$: the flux the $18$ th spectral channel is strongly depleted because of the large coefficients at neighboring wavelengths. 
\item {\it Companion masking in the target image}. One way to to reduce $I_{cross}$ consists on minimizing the magnitude of the $c_k$'s at the location of the companion. Since the off-axis source has already been detected, this can be achieved by masking its pixels in the target image and thus solve the following least-squares problem:
\begin{equation}
\mbox{Find } c_{k} \mbox{ st } min \left \{ ||w_{\mathcal{P}}(x,y) T(x,y) - \sum_{k}^{^{N_{\mathcal{R} }} }  c_k  R_k(x,y) ||_{\mathcal{O} } \right \}
\end{equation}
where $w_{\mathcal{P}} (x,y)$ is equal to $0$ at the location of the detected companion and $1$ everywhere else.  This is equivalent to finding a composite reference which minimizes both the residual in the $\mathcal{O}$ zone, and the impact of the composite reference at the location of the companion:
\begin{equation}
\mbox{Find } c_{k} \mbox{ st } min \left \{ ||T - \sum_{k}^{^{N_{\mathcal{R} }} }  c_k  R_k ||_{\mathcal{O} - \mathcal{P} } + ||\sum_{k}^{^{N_{\mathcal{R} }} }  c_k  R_k ||_{ \mathcal{P} } \right \}.
\end{equation}
This is illustrated on the third panel of Fig.~\ref{Fig::ZonalProcessExplainMasked1}. This approach can be interpreted as a simplified version of d-LOCI, without the coefficient positivity constraint and under the implicit assumption that $\Lambda = 1$. This approach considerably reduces $I_{bias}$ to the point that the flux from the companion is overestimated.
\item {\it Subtraction zone masking and companion masking in the target frame} combines the two approaches and is illustrated on the second panel of Fig.~\ref{Fig::ZonalProcessExplainMasked2}. It does alleviates spectral contamination but overestimates the companion flux, albeit at a much lower level than solely masking the companion in the target image.
\end{itemize} 
The analysis using synthetic companions presented on Fig.~\ref{Fig::ZonalProcessExplainMasked1} and Fig.~\ref{Fig::ZonalProcessExplainMasked2} was carried out at several angular separations and delta magnitudes and consistent results were obtained. While using subtraction zone masking in conjunction with companion masking in the target frame yields good spectral fidelity at low contrast levels, it degrades quickly when the synthetic companions gets fainter.  Replacing the $\mathcal{S}$ zone masking by d-LOCI provides a robust alternative to this problem.
%
\subsection{Results}
\subsubsection{Absolute and relative photometry}
IFS observations of faint companions with instruments such as P1640 yield two astronomical observables: band averaged photometry and low resolution spectroscopy.  Estimating both quantities from aggressively LOCI subtracted data-cubes require two fundamentally different reduction strategies:
\begin{itemize}
\item Band-averaged photometry (J and H bands in the case of P1640): accurate broadband photometry is crucial for deriving effective temperatures and determining masses based on evolutionary models. The band averaged flux is insensitive to the intrinsic spectral shape; this is regardless of  whether or not the companion has absorption or emission bands in the spectrum (main source of errors in $I_{bias}$). Therefore the band-averaged LOCI response is relatively insensitive to underlying spectrum and can be calibrated using synthetic companions with a flat spectra. This is equivalent to work done by several authors \citet{2008Sci...322.1348M,2009ApJ...707L.123T,2011ApJ...729..128C,2011ApJ...728...85J}. LOCI parameters ($N_{A},g,dr,N_{\delta}$) can be tuned to optimize band-averaged performances.  For instance, the bias of the integrated flux over both J and H bands, corresponding to the reductions illustrated on  Fig.~\ref{fig::ExplainExtraction}, can be minimized using a LOCI parameters optimization based on synthetic companions. On this figure, regular LOCI reductions  (middle column of Fig.~\ref{fig::ExplainExtraction}) exhibit more accurate band-averaged photometric estimates in the H band than in the J band: this suggests that the  parameters can be further optimized in  J band.

\item Low resolution spectroscopy: accurate spectroscopy allows further constraints on physical parameters such as effective temperature, surface gravity, and the detection of molecular absorption bands. Here the color response of the algorithm is sensitive to the intrinsic spectral shape (main source of errors in $I_{cross}$). Calibrating this response without foreknowledge of the companion's spectrum is difficult. This is the main problem addressed in this paper: we show here that, even at the detection limit, d-LOCI provides an SED estimates much more accurate than classical LOCI. Importantly, this does not require any calibration of the color response of the algorithm.
\end{itemize}

\begin{figure}
\begin{center}
\includegraphics[width=5in]{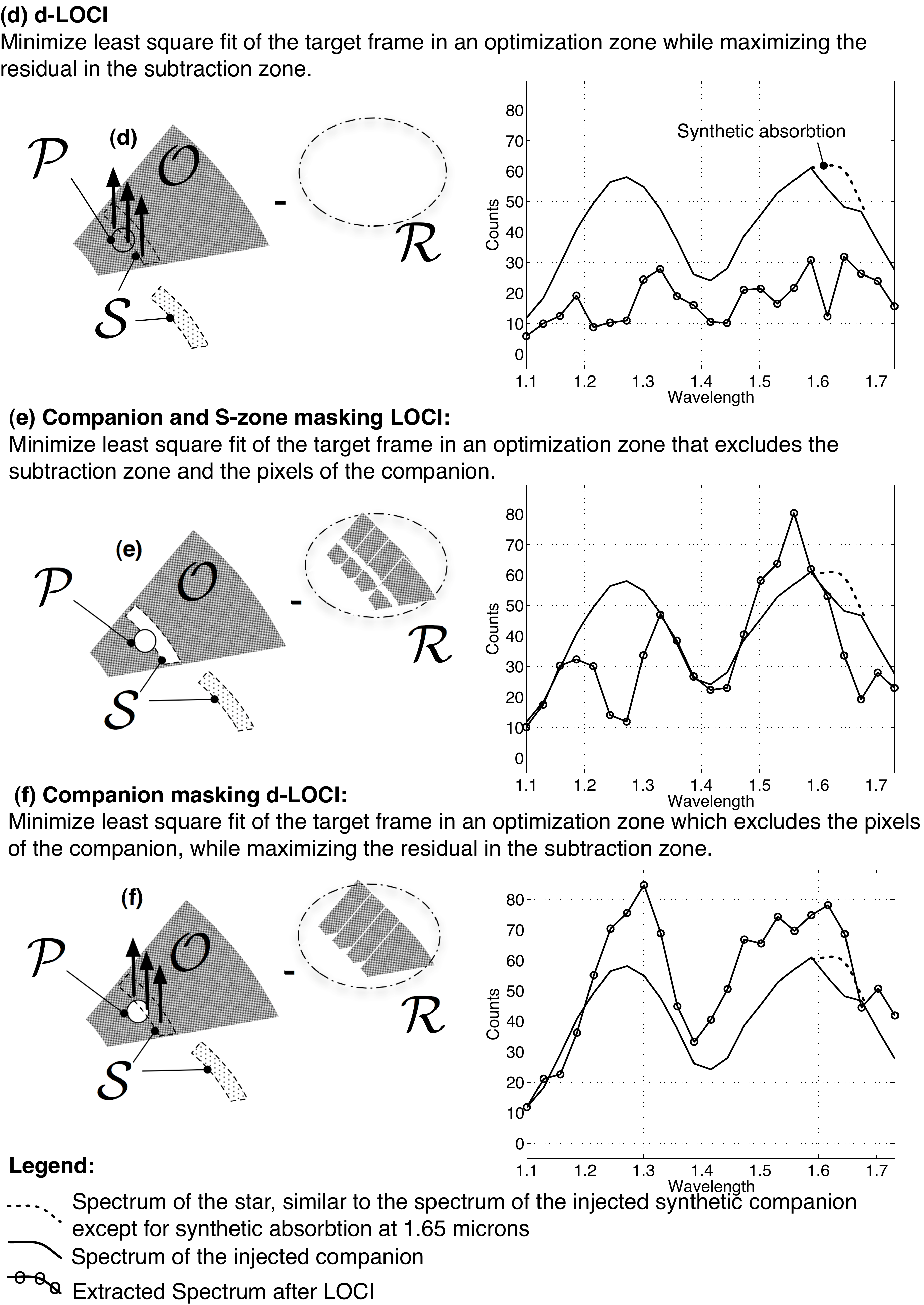}
\caption[Illustration of PSF masking to preserve the flux of the companion]{Illustration of three PSF binary weighting approaches combined with d-LOCI, with their respective performances in the case of a companion slightly fainter than the P1640 speckles, $\Delta m = 6$.  Solutions (e) and (f) provide the best two spectral estimates and their biases are carefully quantified in the present manuscript. $\mathcal{O}$ is the optimizations zone, used to find the best least-squares fit of the speckles realization. $\mathcal{S}$ is the subtraction zone. $\mathcal{P}$ is the position of the planet. $\mathcal{R}$ is the ensemble of reference images. The upwards arrows illustrate the flux preservation constraint in d-LOCI.}
\label{Fig::ZonalProcessExplainMasked2}
\end{center}
\end{figure}

\begin{figure}
\includegraphics[width=5in]{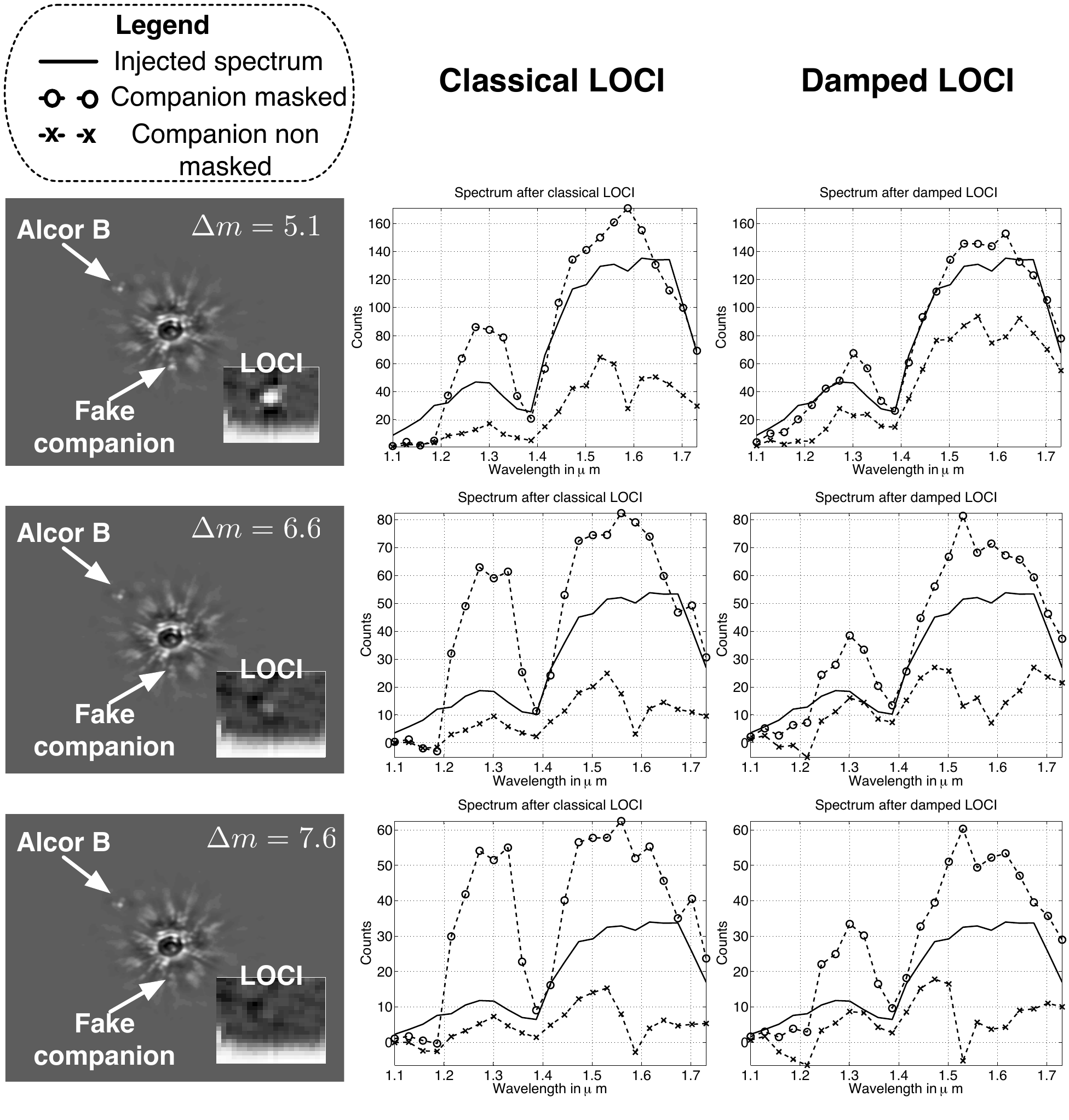}
\caption[Spectral accuracy of the regular and damped LOCI approach]{Qualitative comparison of the spectro-photometric accuracy of different LOCI reductions. We used synthetic companions at $1$ arcsec at three different levels of contrast. The injected spectrum is represented by a solid line in all panels. {\bf Left Panel}: Images before and after LOCI. {\bf Middle panel}: Classical LOCI, solution (a) in Fig.~\ref{Fig::ZonalProcessExplainMasked1} and Fig.~\ref{Fig::ZonalProcessExplainMasked2}, in crossed-dashed line; classical LOCI with dual masking, solution (e), in circled dashed line. {\bf Right panel}: d-LOCI, solution (d), crossed-dashed line, d-LOCI with target frame masking, solution (f), in circled dashed line. {\it When using d-LOCI, solution (f), most of the residual error corresponds to a gray gain, or absolute error. The relative SED error is smaller in this d-LOCI than in the classical masked case, solution (d)}. This figure illustrates approaches which do not include any iterations on the underlying companion's spectrum or LOCI parameters}
\label{fig::ExplainExtraction}
\end{figure}

We first discuss qualitatively the performance of four possibilities, with masking/no-masking and with regular/damped LOCI, with respect to band-averaged photometry and low resolution spectroscopy.  To do so, we introduced a synthetic companion located at $\simeq 1''$ separation in $4$ data-cubes on the star Alcor, \citet{2010ApJ...709..733Z}. This is about $10$ minutes of total exposure, with each 3D cube composed of $N_{\lambda} = 23$ wavelength frames. The 4 data-cubes allow a reference frame set of 80 frames. Unlike the A5V host Alcor,  the 1.2 - 1.8 micron SED of the synthetic companion has H-band excess; this lowers the companion signal-to-noise at short wavelengths (J-band) where the surrounding speckles are brighter.  We then ran all reductions with constant LOCI parameters $dr=2$, $N_{\delta} = 1$, $g = 3$ and $N_{A}=100$; this choice of $N_{A}$ leads to solving a set of inverse problems that are slightly over-constrained, with 80 degrees of freedom and 100 constraints. The results are shown in Fig.~\ref{fig::ExplainExtraction}, for three different companion magnitudes ranging a factor of $10$ in flux. The leftmost column shows the before and after LOCI images used for these analyses. The ``post-reduction'' inset illustrates the companion's detection level in a single spectral channel. These images were obtained with classical LOCI and no priors on the companion's location. The behavior of the four different reductions illustrated in this figure allow insights into the algorithmic photometric and spectro-photometric accuracy of both unmasked and masked approaches. We now refer the reader to  Fig.~\ref{Fig::ZonalProcessExplainMasked1}, Fig.~\ref{Fig::ZonalProcessExplainMasked2} and Fig.~\ref{fig::ExplainExtraction}:

{\bf Flux preservation properties of unmasked LOCI}: The extracted spectral profiles represented by the lines labelled ``Companion non masked'' in Fig.~\ref{fig::ExplainExtraction} correspond to the implementation of non-masked reductions with classical LOCI (center column), solution (a), top panel of Fig.~\ref{Fig::ZonalProcessExplainMasked1} and with d-LOCI (right column), solution (d), top panel of Fig.~\ref{Fig::ZonalProcessExplainMasked2}. d-LOCI exhibits an improvement in
the algorithmic throughput of a factor of two. As there is no prior about the companion's location, and as the SNR of both solutions are similar (\S 3), we find that d-LOCI preserves a larger portion of the companion's  flux. Moreover the top set of panels of Fig.~\ref{fig::ExplainExtraction} demonstrate that classical LOCI without binary weighting can lead to gray spectral transmitivities. This holds when the companion is as bright as the speckles as is seen in \citet{2010arXiv1008.4582B}, where the authors used a set of synthetic companions to derive a gray algorithmic spectral response for HR8799 b. However this property quickly degrades with contrast and more robust methods are needed when the companion is fainter than the speckles.

{\bf Spectro-photometric accuracy  in masked or damped approaches}: in the middle column of Fig.~\ref{fig::ExplainExtraction}, the spectral profiles represented by the lines labelled ``Companion masked'' correspond to the implementation of  LOCI with both subtraction zone masking and companion masking in the target frames. This refers to solution  (e) in Fig.~\ref{Fig::ZonalProcessExplainMasked2}. In the right panel the same spectra are generated by d-LOCI with companion masking in the target frames. This refers to solution (f) in Fig.~\ref{Fig::ZonalProcessExplainMasked2}. The performance of both methods degrades as the companion gets fainter. However, the former clearly results in larger spectro-photometric errors than the latter (with damping).

These findings are summarized in Table.~\ref{tab::algocomparison}, which describes the various levels of refinement of LOCI reductions necessary to carry out accurate measurement of band-averaged photometry and low resolution spectroscopy. We identified three levels of synthetic companion brightness: brighter than speckles (can be detected in a broadband band image but its photometry is contaminated by the speckles ), as bright as the speckles (can only be identified by its immobility in a movie of sequential wavelengths), fainter than the speckles. The level of refinement of LOCI reduction was qualitatively inferred from the series reductions carried out to generate Fig.~\ref{fig::ExplainExtraction} ($15$ different companion brightness ranging from $\Delta m = 5$ to $\Delta m =8$). Note that in this paper, no effort was made to optimize the sensitivity to faint off-axis sources. For details pertaining to the behavior of the detectability of P1640 as a function of observing conditions (e.g seeing, host star magnitude, integration) and reduction procedure (e.g optimization of LOCI parameters), we direct the reader to \citet{JustinSpeckle}. Next we quantify the absolute and relative spectral accuracies that are obtained using the two most promising approaches identified in Table.~\ref{tab::algocomparison}: dual PSF masking with classical LOCI and single PSF masking with d-LOCI.
\subsubsection{Band-averaged photometry absolute error}
\begin{figure}
\includegraphics[width=5in]{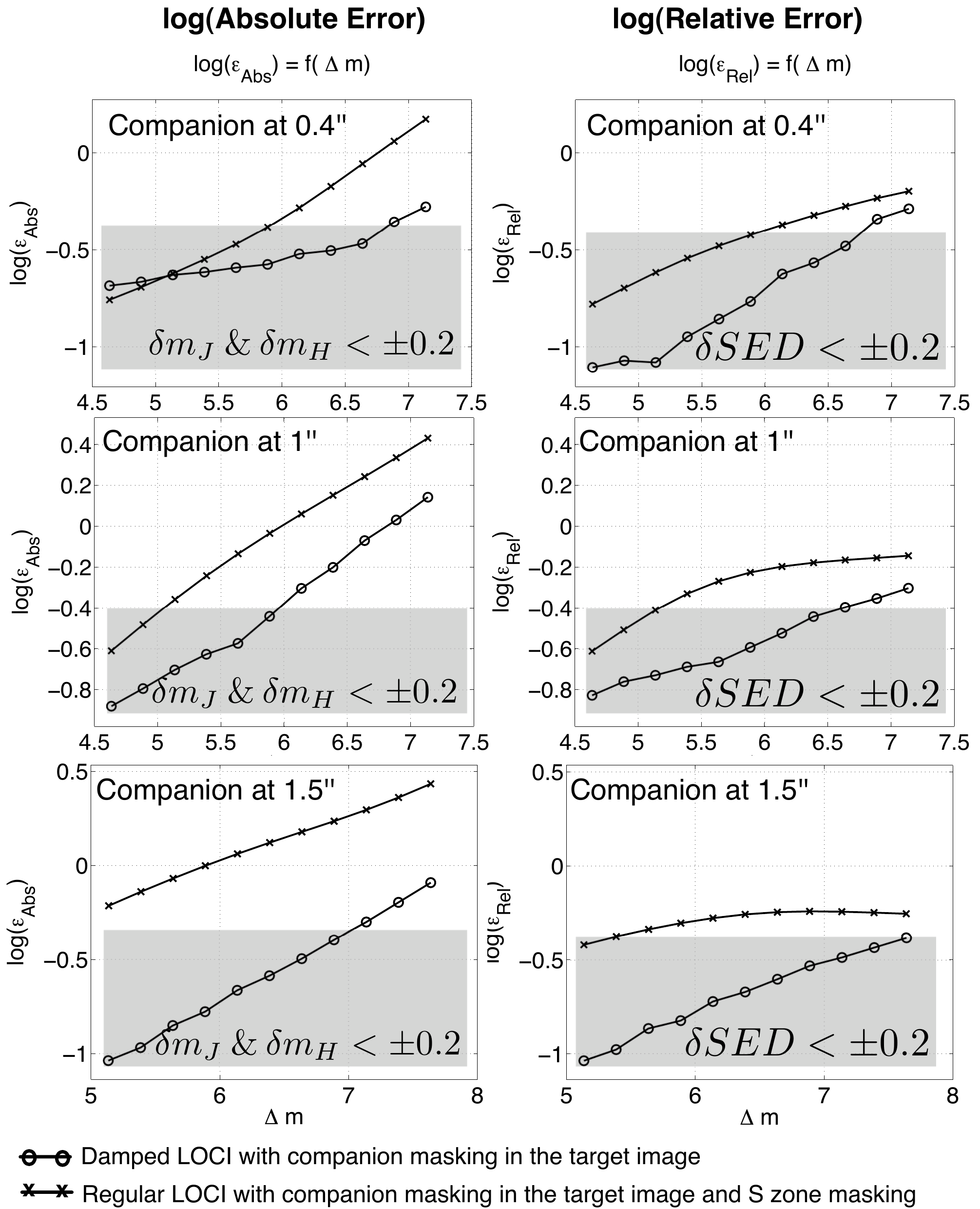}
\caption[Log normalized error in the spectrum estimation as a function of delta magnitude]{Quantitative comparison of the spectro-photometric accuracy for LOCI reductions (d) and (e) as a function of contrast at three different angular separations. Left: band-averaged photometric error.  Right: bias in the SED estimate. The gray area corresponds to published band-averaged photometric error bars on recently discovered faint companions using ADI: $\delta m_{J} \: \& \: \delta m_{H} < \pm 0.2$ mags.{\it Even in the challenging ``IFS +no field rotation'' configuration, d-LOCI delivers systematic errors on the SED of the order of state of the art uncertainties derived from band-averaged ADI observations, this all the way to the detection limit of the P1640 instrument.}}
\label{fig::QuantifyExtraction}
\end{figure}

Here the absolute spectro-photometric error is the root mean squared error in the measurement of the companion flux, integrated over the spectral channels spreading the J and H band. 
\begin{equation}
\epsilon_{Abs} = \frac{\sqrt{\sum_{p=1}^{N_{\lambda}} (s^{telluric}_{p})^2 (I_{p}^{C} - I_{p}^{C,LOCI})^2 }}{\sum_{p=1}^{N_{\lambda}}  s^{telluric}_{p} I_{p}^{C}}
\end{equation}
where $s^{telluric}_{p}$ is the spectral response function, derived using reference A stars according to the method detailed in \citet{NeilPipeline}. The left column of  Fig.~\ref{fig::QuantifyExtraction} shows how this metric varies as a function of the brightness and angular separation of a potential companion. The band-averaged photometric accuracy scales with the  absolute spectro-photometric error. For comparison, the average level of  photometric uncertainty in current ADI broadband discoveries of faint companions ($\delta m \simeq 0.2$  e.g. \citet{2008Sci...322.1348M,2009ApJ...707L.123T,2011ApJ...729..128C,2011ApJ...728...85J}) is indicated on Fig.~\ref{fig::QuantifyExtraction}. d-LOCI consistently leads to smaller absolute spectro-photometric errors for the reduction parameters ($N_{A},g,N_{\delta},dr$) tested here. The improvement in $\epsilon_{Abs}$ between dual masked LOCI and d-LOCI varies as a function of angular separation. This suggests, as discussed above, that the LOCI parameterization can be further optimized in order reduce the systematic error bar on band-averaged photometry.  Note that most of the spectro-photometric error when using d-LOCI in the faintest case of Fig.~\ref{fig::ExplainExtraction} (bottom-right panel) consists of a band-average bias. Such an error can be mitigated using a set of synthetic companions to optimize LOCI parameters. In the faintest case, the actual improvement in relative SED (up to a scaling constant) provided by d-LOCI, when compared to classical LOCI (bottom-middle panel of Fig.~\ref{fig::ExplainExtraction}), is actually significant. We quantify this improvement in the next sub-section.

Note that the two reductions methods considered on Fig.~\ref{fig::QuantifyExtraction} assume that a companion has already been detected and has been masked in the target frames. As a consequence they can only be carried out for sources already identified using non-masked LOCI reductions. We separately evaluated the detectability of the synthetic companions in Fig.~\ref{fig::QuantifyExtraction} and did not include sources below the band-averaged $5-\sigma$ level limit. We find that in the case of the P1640 data-set used for these calculations, this regime occurs at $\Delta m \simeq 7.3$ with slight variations depending on the angular separation. While this number is relatively modest we emphasize that this is mostly due to our conservative choices of data-sets (average observing conditions, no wavefront calibration before observations) of fake spectrum (faint in the J band), short exposure times  ($9$ minutes) and reduction parameters ($N_{A},g,N_{\delta},dr$). Further optimizations of theses parameters can surely lead to deeper detectability and absolute error that is consistent with the band averaged photometric error bars published in previous ADI discoveries; all the way to the $5 -\sigma$ detection limit. 

%
\subsubsection{Low resolution spectroscopy: relative error}
When seeking to characterize the spectral profile of a discovered companion within its near infrared bands, calibrating the color response of the LOCI reduction without foreknowledge of the companion's spectrum is difficult. When it is sufficiently bright with respect to the speckles, the special case of a gray spectral response for the reduction arises. \citet{2010arXiv1008.4582B} provides a thorough illustration of this regime in the case of HR8799b.  When the algorithmic color response is gray, it is safe to assume that $I_{bias}$ is responsible for most of the flux depletion. Indeed, since $I_{cross}$ mostly amplifies local absorption features,  it is unlikely to be constant over the full spectrum. In this particular case both the photometry and the low resolution spectral features of the object can be estimated. However, as the companion gets fainter, the reduction's spectral response starts varying with wavelength. Without prior knowledge about the spectral profile of the companion, it is very difficult to disentangle whether such variations are the sole result of LOCI (e.g $I_{bias}$) or stem from the spectrum of the faint source propagated though LOCI (e.g $I_{cross}$). One solution to address this issue would be to carry out a series of iterative reductions: (1) extract a biased spectrum from a preliminary LOCI reduction, (2) subtract this spectrum from the non-reduced data, (3) run a secondary LOCI reduction, (4) iterate. Since the biases depend on the set of LOCI parameters chosen for the reduction ($N_{A},g,N_{\delta},dr$),  this process should be repeated over a wide set of parameters in order to ensure that the biases on the SED are minimized. In this paper we answer an orthogonal question. Namely, {\it how to minimize the relative spectro-photometric bias in one iteration, for a well chosen but not optimized set of LOCI parameters?}

We quantify the relative spectro-photometric error as measure of the uncertainty in estimating the Spectral Energy Distribution of the discovered object: 
\begin{equation}
\epsilon_{Rel} = \sqrt{N_{\lambda}\sum_{p=1}^{N_{\lambda}}(s^{telluric}_{p})^2 \left (\frac{I_{p}^{C,LOCI}}{\sum_{k=1}^{N_{\lambda}} s^{telluric}_{p} I_{k}^{C}} - \frac{I_{p}^{C}}{\sum_{k=1}^{N_{\lambda}} s^{telluric}_{p} I_{k}^{C}}\right )^2 }.
\end{equation} 
In other words $\epsilon_{Rel}$ captures of how well the overall shape of the low resolution spectrum can be retrieved, up to a scaling constant. Assuming that the band-averaged photometry of the object has already been estimated, this SED provides the quantitative basis to further characterize the discovered companion. In particular it provides remarkable level arm to estimate both surface gravity, atmospheric chemistry  \citet{2006ApJ...639.1095B,1995Sci...270.1478O} and, to a lesser extent, the atmospheric cloud content of the faint object \citet{2010ApJ...725.1405B}. Therefore the spectro-photometric error will have a direct impact on the estimation of the physical parameters of the faint exo-planets discovered by campaigns such as GPI, SPHERE, P3000-P1640. Without any priors about the spectral content of such objects, calibration of the biases introduced by the reduction using synthetic sources becomes a very delicate exercise. A first avenue that can be explored when seeking to address this issue of faint companion spectral characterization is to devise a new class of algorithms that rely on priors about the speckle field (e.g. \citet{Burke:10,2011arXiv1103.4766M}) or the spectral type of the companion (e.g. spectral deconvolution as in \citet{2002ApJ...578..543S}). The solution proposed in the present manuscript, d-LOCI, focuses class of solution which assumes no priors about the quasi-static errors or the companion.

The right panel of Fig.~\ref{fig::QuantifyExtraction} demonstrates that  d-LOCI is a robust approach when seeking to characterize the spectral features of a faint companion without any foreknowledge of its spectral type. Indeed, the relative error using d-LOCI in conjunction with target frame masking is, for most separations and levels of contrast, half an order of magnitude smaller than the one obtained using dual masked classical LOCI.  More importantly it yields SED uncertainties below the error-bar level of state of the art band-averaged photometric estimates (\citet{2008Sci...322.1348M,2009ApJ...707L.123T,2009ApJ...705L.204M}). This feature is quite remarkable since it reduces the spectro-photometric uncertainties associated with observations with an IFS {\it without} field rotations to the level of  observations carried out in a more favorable mode: ADI  ($I_{cross} = 0 $) and images integrated over several P1640 channels (more photons from the companion). For the brightest synthetic companions tested here, the SED bias  obtained using d-LOCI is of the order of the P1640 telluric calibration uncertainties. We also emphasize that such results are obtained without any iterations on LOCI parameter or underlying companion spectrum, and thus d-LOCI reduced images are a good first guess when seeking to obtain even higher spectro-photometric accuracy using iterative schemes.

 d-LOCI is thus an efficient method when one seeks to estimate the SED of a detected point source in P1640 data, without any prior knowledge about the spectral profile of the companion or about the chromaticity of the post coronagraphic speckles.  This result ought to be carefully extrapolated to other instruments, in particular using  similar tests based on synthetic companions. We expect however that such conclusions will stand when the raw contrast of direct imaging campaigns will improve, either using a real-time calibration system (\cite{wallace:74400S}) or advanced pre-observing calibration methods (\cite{2010Natur.464.1018S}). We thus envision that the approach suggested in herein will be critical to the estimation of the physical parameters and atmospheric content of exo-planets detected via future direct imaging campaigns \citet{2007arXiv0704.1454G,2006SPIE.6269E..24D,dekany:62720G,2011PASP..123...74H}.
\section{Conclusion}

In this paper we discuss an algorithm for Point Spread Function subtraction in data acquired with an Integral Field Spectrograph, that is efficient and accurate for retrieving the spectral content of faint companions located within several diffraction widths of a target star.  
A previous communication by \citet{JustinSpeckle} used the newly functional Project 1640 spectrograph, \citet{2011PASP..123...74H}, as a laboratory to devise a robust method purely for off-axis faint source detection. This method relies on incorporating the spectral diversity inherent in IFS data into a Locally Optimized Combination of Images (LOCI) PSF calibration scheme. In that paper, we reported H-band $5 \;\sigma$ detection limits of $ \simeq 10^{-5}$ in contrast, at 1 arcsec separation, in a $20$ min of exposure on an $H = 3$ star. In the present paper, we expand upon the general LOCI scheme, and discuss the spectral characterization of already detected faint companions. We show that when LOCI is tuned to generate an {\it aggressive} composite calibration PSF, the spectro-photometric content of the faint object is biased. It is affected by both signal depletion, which is common to all LOCI implementations, and by spectral cross-talk with neighboring wavelengths, which is specific to IFS imaging spectroscopy. Herein, we introduced a new algorithm, ``damped LOCI'' (d-LOCI) that alleviates many of these shortcomings.

We modified the formalism underlying LOCI PSF calibration by introducing an augmented cost function that seeks to preserve the flux from the companion in question. We compared the efficiency of LOCI and d-LOCI for ``detectability'', and found that both approaches performed similarly in the various test cases. We identified second order trends in the behavior of SNR as a function of exposure time and LOCI parameters. Such trends can be explained by the conditioning of the inverse problem that is at the heart of LOCI reduction, and have only marginal bearing on the conclusions pertaining to ``detectability'' discussed in \citet{JustinSpeckle}. While d-LOCI clearly increases the signal of faint companions, it results in a noise floor that is shallower than that obtained in LOCI, and thus the detectability metric of both approaches is similar.

We finally delved into the problem of spectral extraction. We first used synthetic companions to explore the behavior of classical LOCI reductions, implemented with a variety of binary weightings. Our tests showed that a classical implementation of LOCI, with well chosen parameters as in \citet{2007ApJ...660..770L}, can be applied to companions slightly brighter than the speckles and yields an ``almost gray'' algorithmic throughput that yields a 10 percent  error in SED estimates. In this regime we find the spectral fidelity can be improved by the use of a dual binary weighting when posing the PSF reconstruction problem. Since we expect telluric calibration errors below $10$ percent, the main source of errors in the spectro-photometric characterization of the discovered companions resides in bias within the reduction algorithm. Moreover these biases increase rapidly as function of contrast, and when the companion is fainter than the speckles, the error in the SED estimate can be as large as 50 percents. Using d-LOCI, in conjunction with a binary weighting of the target frames at the location of the companion, consistently reduces this error by a factor of five: it lowers the algorithmic errors in the SED close to the levels imposed by the telluric calibration. Translating these uncertainties in error bars on the estimates of the physical parameters of potentially discovered companions is beyond the scope of this paper, and will be illustrated in a future communication. All results are obtained using the Project 1640 IFS data, but we fully expect that these results can be generalized to future high contrast imaging instruments operating in conjunction with advanced wavefront calibration techniques. These next generation instruments (\citet{2007arXiv0704.1454G,2006SPIE.6269E..24D,dekany:62720G,2011PASP..123...74H}) will operate at contrasts significantly higher than those possible with the current PALAO+P1640 configuration, and they will use Integral Field Spectrographs as science cameras.  Accurate in-situ spectral characterization of faint exo-planets buried in the residual speckle fields of the long exposures will thus be a crucial component of upcoming campaigns. It will lead to far tighter constraints on planetary bulk physical properties and atmospheric chemistry and would be a direct application of the d-LOCI methods discussed herein.

\clearpage

\clearpage

\clearpage

\clearpage

\clearpage

\begin{table}
 \caption{SUMMARY OF OUR FINDINGS} 
\begin{center}
\begin{tabular}{m{4cm}m{4 cm}m{4cm}m{4cm}}
\hline 
\; \; & \multicolumn{3}{c}{Preferred LOCI method $^{(a)}$for given companion brightness and science goals}\\ 
 \hline
 Companion brightness   &  Photometry w/o iterations$^{(b)}$ &    Photometry with iterations$^{(c)}$  & SED estimates$^{(d)}$ \\ 
 \hline
 \hline
Brighter than speckles  &  dual masked LOCI &  LOCI & dual masked LOCI  \\
 \hline
As bright as speckles  &  d-LOCI &  dual masked LOCI & d-LOCI  \\
 \hline
Fainter than speckles  &  NONE &  d-LOCI & d-LOCI  \\
 \hline
\end{tabular}
 \label{tab::algocomparison}
 \note{(a)LOCI reduction with the smallest degree of sophistication that delivers a given science goal for a given companion brightness. \\(b) The iterations can be carried out on the LOCI parameters or the brightness of the object. In any case a sequence of synthetic companions injections and LOCI reductions is necessary in this scheme. \\(c) e.g one single LOCI reduction is sufficient to provide a robust photometric estimate.\\(d) Here we assume that  LOCI biases on the SED depend on the underlying spectrum, and thus are very difficult to calibrate using iterations without any prior on the faint companion.}
\end{center}
\end{table}

\clearpage

\appendix

\section{Appendix: Frame selection and geometry}
\begin{figure}
\includegraphics[width=6in]{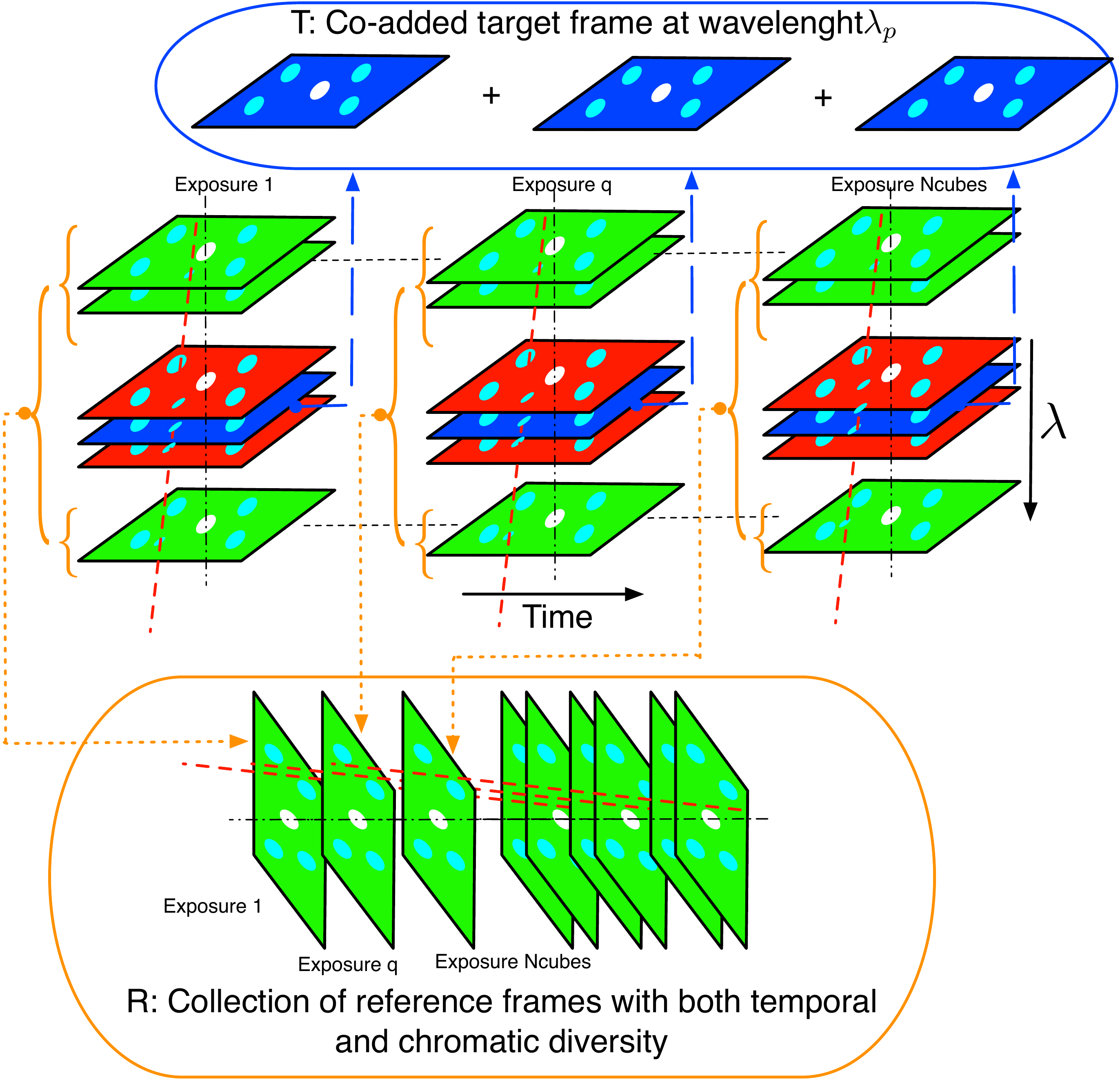}
\caption[Illustration of the frame classification method]{Illustration of the frame classification method. The data consists in a collection of cubes, represented in the middle panel. Each cube corresponds to one exposure and each slice within a cube is an image at a given wavelength: here, the temporal dimension is represented horizontally and the wavelength vertically. We co-add the frames at the wavelength $\lambda_{p_0}$ into a target frame T,  top panel. We create a library of references frames that will be used to build the composite calibrator $\{ R_{k} \}$. These references are chosen so that the flux of a potential companion at neighboring wavelengths is minimal at the radial location of interest (e.g. minimize self-subtraction). The reference ensemble presents both chromatic and temporal diversity.}
\label{Fig::TowerSortIllustration}
\end{figure}

\begin{figure}
\includegraphics[width=6in]{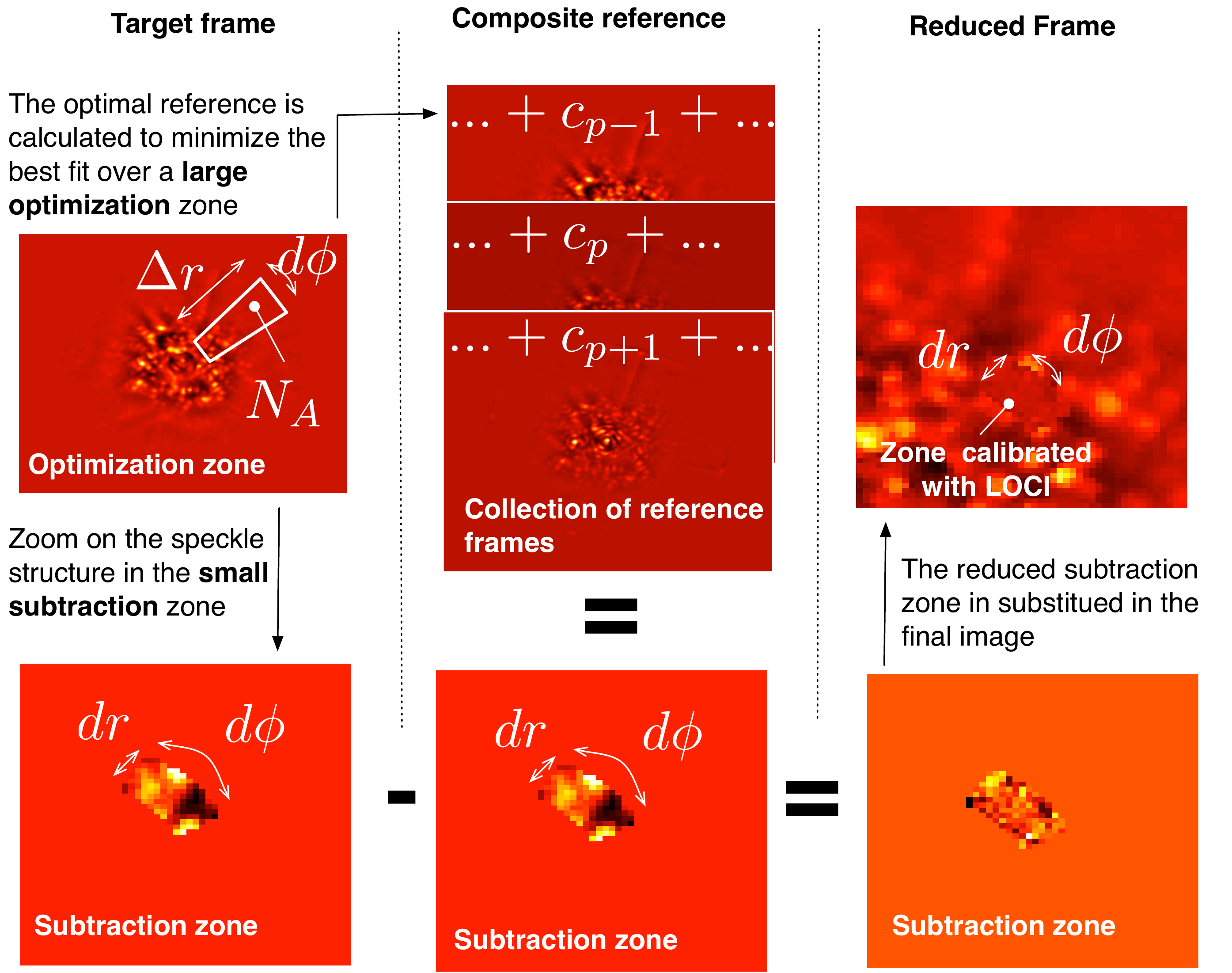}
\caption[Illustration of the optimization and subtraction zones associated with LOCI]{Geometry associated with a polar implementation of LOCI. The reference coefficients are chosen in order to obtain the best least-squares fit of the target speckle pattern in the {\it large} optimization zone, $\mathcal{O}$. The corresponding composite reference PSF is then subtracted from the target frame only in a {\it smaller} area, the subtraction zone $\mathcal{S}$, that is included in $\mathcal{O}$. A zoom on the $\mathcal{S}$ after LOCI subtraction is displayed in the rightmost panel of the figure.}
\label{Fig::ZonalProcessExplain}
\end{figure}

In this communication we consider the case of a series of IFS cubes that have already been centered and spatially re-scaled so that the optical artifacts are stationary within the image space and the astronomical image has been stretched radially, \citet{JustinSpeckle,2002ApJ...578..543S}. Proper scaling and centering of the images is crucial to subsequent data analysis. We carry out this stage by finding the centering and scaling parameters that maximize the auto-correlation between two images. We solve this non-linear search using the sub-pixel image registration algorithm described by \citet{Guizar-Sicairos:08}, which has been modified using the Discrete Fourier Transform algorithm described in \citet{2007arXiv0711.0368S}, in order to accommodate for the scaling parameter. The series of images resulting from such a calibration corresponds to a virtual four dimensional cube, with two spatial dimensions, one spectral dimension and a temporal dimension. We seek to devise a reduction algorithm that will stack the signal of a potential companion for a given wavelength, while subtracting the artifacts due to optical imperfections. We write each frame  in our four dimensional cube as $I_{p}^{(q)}$ where the subscript $p = 1 \; ... \; N_{\lambda}$ stands for the spectral channel, and the superscript   $q = \; 1 ... N_{cubes}$ stands for the cube number, or the temporal dimension. Associated with these slices is  a collection of scaling parameters addressed similarly, $\gamma_{p}^{(q)}$. We first build an integrated target frame $T$, with an associated scaling parameter $\gamma_{p_0}$, that corresponds to the sum of the frames at the wavelength of interest $\lambda_{p_0}$:
\begin{eqnarray}
T &=& \sum_{q = 1}^{N_{cubes}}I_{p_0}^{(q)}\\
\gamma_{p_0}& =& \frac{1}{N_{cubes}}\sum_{q = 1}^{N_{cubes}}\gamma_{p_0}^{(q)}
\end{eqnarray}
This is illustrated in the top panel of Fig.~\ref{Fig::TowerSortIllustration}, where slices of the same wavelength are co-added. We seek to create a composite reference that locally subtracts the quasi-static speckles without removing the flux from a putative companion. Thus, at a given radial location within the target frame, written as $r$, we need to sort the remaining frames into two subsets: $\{ R_{k} \}$, the subset of slices that will be used to create the composite reference and $\{ J_{k} \}$, a subset of ``useless''  frames whose wavelength is too close to $\lambda_{p}$ so that the image of a potential companion would overlap with the region of interest. Since the several wavelengths in the data-cube provide radial diversity, the image of a potential companion moves radially in the rescaled space, we can rewrite the self-subtraction condition established by \citet{2007ApJ...660..770L} as:
\begin{eqnarray}
\mathcal{R} &=& \{ R_{k} \} = \{ I_{p}^{(q)}: \;|r (\gamma_{p}^{(q)} -\gamma_{p_0} )| \geqslant N_{\delta}  W\} \\
\mathcal{J} &=& \{ J_{k} \} = \{ I_{p}^{(q)}: \;|r (\gamma_{p}^{(q)} -\gamma_{p_0} )| < N_{\delta}  W\}
\end{eqnarray}
where $W$ is the Full Width Half Maximum (FWHM) of a PSF at wavelength $\lambda_{p_0}$, and $N_{\delta}$ is a parameter that determines the aggressiveness of  the LOCI reduction algorithm. This sorting process involves rejecting the slices whose wavelength is too close to $\lambda_{p_0}$ from the collection of references, so that there is minimal companion flux in the composite references. We actually do not integrate over time the slices  in the {\it reference ensemble} and we let LOCI choose the best coefficients to form an optimal reference PSF over time and wavelength. The strategy summarized on Fig.~\ref{Fig::TowerSortIllustration} was chosen on order to mitigate computational cost. Indeed, calculating a LOCI reference PSF for each slice of each exposure using PSFs at other wavelengths and slightly de-correlated in time, would mean running $N_{cubes} \times N_{\lambda}$ LOCI reductions, with each reduction looking for coefficients over an ensemble of  $N_{cubes} \times (1 - \epsilon (N_{\delta})) N_{\lambda}$ reference PSFs ($\epsilon(N_{\delta})$ is the number of slices excluded to avoid cross talk). Instead our reduction strategy only leads to $N_{\lambda}$ LOCI reductions with each reduction looking for coefficients over an ensemble of  $N_{cubes} \times (1 - \epsilon (N_{\delta})) N_{\lambda}$ reference PSFs. The performances of each reductions strategy in terms of detectability depends on the actual shot noise level of the speckle field (e.g. whether LOCI actually reconstructs the true realization of the speckle field in the target image, or fits the photon noise associated with it). The true merit of a  given frame selection strategy thus depends strongly on observational conditions such as stellar magnitude or integration time. Finding the optimal  binning scheme for given observational conditions is beyond the scope of the present manuscript and we did not delve into such trade-offs. Instead, we chose a hybrid frame selection approach which preserves the temporal information in the ensemble of reference PSFs while integrating the time variability of the target science image. The next step of the reduction algorithm is to find a series of optimal coefficients in order to build the composite reference. This is the stage that can lead to a loss of the spectral information of the detected companion. In order to formally describe the challenges of LOCI reductions we first briefly review the notations involved in the framework introduced by \citet{2007ApJ...660..770L}.
As shown on Fig.~2, the local composite reference PSF is constructed by finding a linear combination of reference frames, $\{R_{k} \in \mathcal{R} \}$, that yields the best least-squares fit of the speckle pattern in the target image, $T$. This locally optimal calibration image is then subtracted from the target image. However, in order for this reference to only describe the speckle field and not the flux of a potential companion, its coefficients ought to be determined in a zone of the image that is larger than the zone within which the subtraction will occur.  Thus the LOCI algorithm uses two regions of interest in the images, a subtraction zone, $\mathcal{S}$, where we will subtract the composite reference from the target, and a large optimization zone, $\mathcal{O}$, that surrounds $\mathcal{S}$, that we will use to find the weights of the composite reference. We follow the presentation of \citet{2007ApJ...660..770L} and use an annular geometry, as illustrated in Fig.~\ref{Fig::ZonalProcessExplain}. The optimization zone is defined by its area, expressed in number of PSF cores, $N_{A}$,  and its aspect ratio $g$. The aspect ratio is a unit-less quantity that scales with the radial elongation of the zone $\mathcal{O}$,
\begin{equation}
g = \frac{\Delta r}{(r + \Delta r/2) \Delta \phi}
\end{equation}
where $r$ is the radial location of the inner annulus of the optimization zone, $\Delta r$ its radial extent and $\Delta \phi$ its azimuthal extent. They can be written as a function of $N_{A}$, $g$ and $W$:
\begin{eqnarray}
\Delta r &=& \sqrt{\frac{\pi N_{A} g W^2}{4}}\\
\Delta \phi &=& \left(\frac{g}{2} + \frac{2 r }{W} (\frac{g}{\pi N_A})^{1/2} \right)
\end{eqnarray}
In the geometry chosen here, illustrated in Fig.~\ref{Fig::ZonalProcessExplain}, the inner radius of the subtraction and optimization zones is similar and equal to $r$. Their azimuthal span is likewise identical and equal to $\Delta \phi$. The radial extend of $\mathcal{S}$ zone is $dr$  while the $\mathcal{O}$ zone stretches all the way to $r + \Delta r$. The efficiency of the LOCI reduction strongly depends on the choice of the parameters $N_{A}$, $g$, $dr$ and $N_{\delta}$. \citet{2007ApJ...660..770L} explored the parameter space for LOCI applied to ADI, and \citet{JustinSpeckle} presented similar work in the case of chromatic differential imaging. In this paper we show that a standard LOCI implementation applied to IFS data yields very good results in term of detectability, but perturbs the spectrum of the companions discovered. While the errors associated with the spectral estimation can be reduced by exploring the $N_{A}, g,dr,N_{\delta}$ parameter space, we propose a more direct solution that consists of constraining the problem associated with the LOCI coefficient search.

\section{Appendix: Detection and Signal to Noise Ratio}
\begin{figure}
\includegraphics[width=6in]{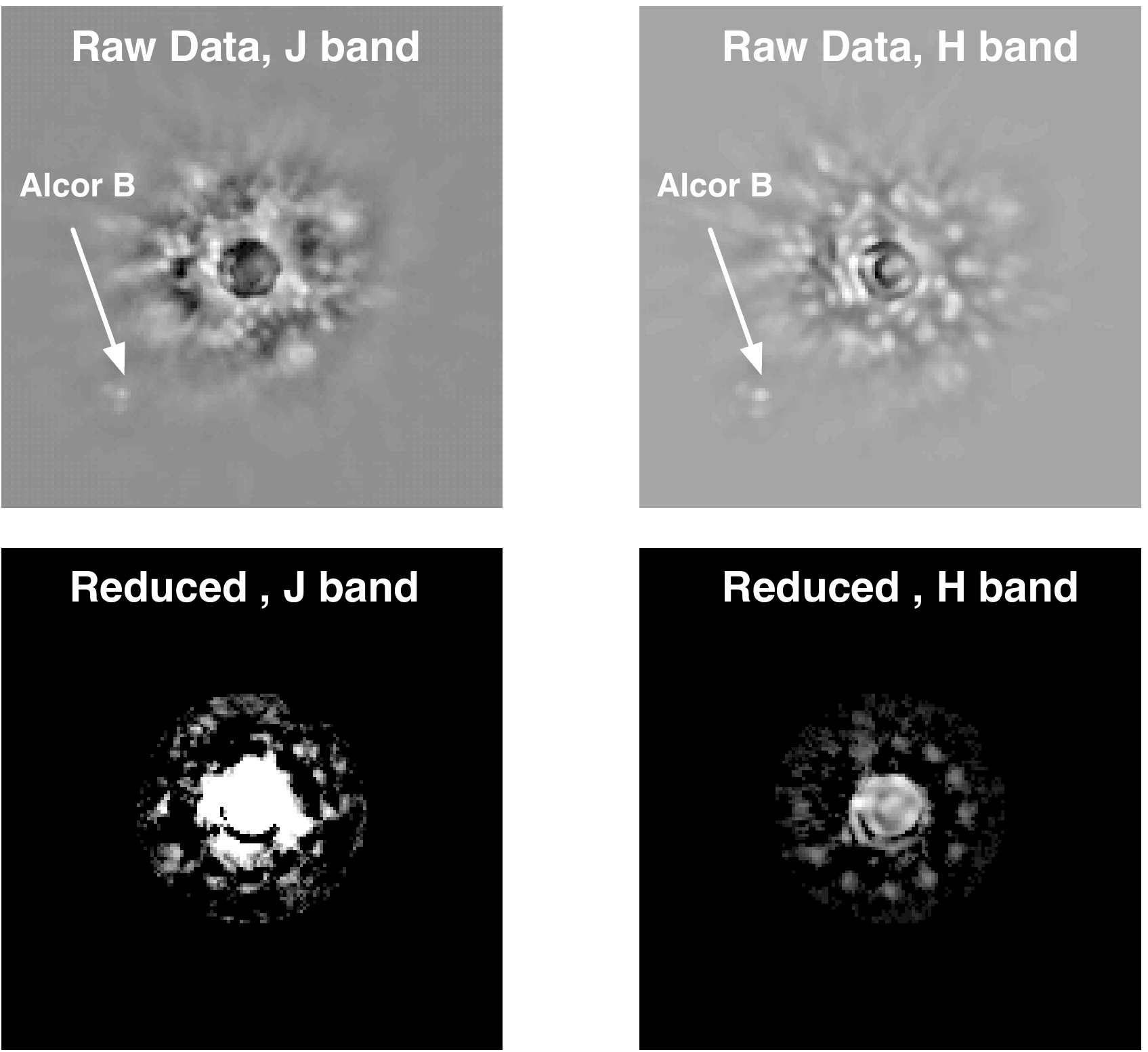}
\caption[Illustration of the data set used for our faint companion detection test]{Data set used for SNR calculations. Left Column: Raw (top) and processed (bottom) images in the J band. Right Column: Raw (top) and processed (bottom) images in the H band. We used a spiral of $11$ synthetic companions with a flat spectrum whose angular separation vary from $0.3$ to $0.5$ arcsec. Because of the better performance of the Adaptive Optics system at longer wavelengths, the raw contrast is more favorable in the H-band than in J-band. In this example the exposure time is $140$ s. Since the average contrast in J-band is $\simeq 10^{-4}$ and $\simeq 10^{-3}$ in H-band the synthetic companions are detected with a higher signal to noise in the latter case. This figure was generated using images reduced by d-LOCI.}
\label{fig::SummaryFakePlanets}
\end{figure}

%
We first study the impact of  d-LOCI on the detectability of faint point sources and show that the damping introduced herein only marginally affects the SNR of detected companions. To do so, we use a data set obtained on the bright A5V star ($V = 4$) Alcor on March $16$ $2009$. We introduced in the data a spiral of $11$ synthetic companions with a flat spectrum, at angular separations varying from $0.3$ to $0.5$ arcsec.  The injection of the artificial sources occurs as high up as possible in the cube processing pipeline, as described in \citet{JustinSpeckle}. Alcor bright visible magnitude enables good performances from the Palomar Adaptive Optics system (PALAO, \cite{dekany:62720G}). A faint M-dwarf was recently detected using the P1640 spectrograph, \citet{2010ApJ...709..733Z}, making this data set an ideal test case. The contrast  of the astronomical companion is moderate, $1/30$ as estimated by \citet{2010AJ....139..919M}, and the secondary star is visible above the speckle halo: it provides a second reference for the photometry of the synthetic companions inserted. Fig.~\ref{fig::SummaryFakePlanets} presents unprocessed images and the results of a data reduction, both integrated in J and H-bands. Since the speckles are brighter at shorter wavelengths, and the spectrum of the synthetic sources injected is gray, the overall contrast associated with them varies between $10^{-2}$ and $10^{-3}$ in the H band, and $10^{-3}$ and $10^{-4}$ in the J band. As a consequence some companions are too faint to be detected in the J reduced images. These results were obtained using the the following parameters: one exposure of $140$ seconds, $N_{\lambda} = 23$ slices across the cube of interest, $N_{A} = 100$, $g = 1$, $dr = 2$ and $N_{\delta} = 1$. This choice of parameters,  in particular $g =1$, implies that two adjacent synthetic companions might be located in the same LOCI zones, and thus might artificially hamper the SNR of reduced images. However we included our ``conservative'', $g = 1$, results in this manuscript, as its purpose was not do quantify the ultimate detection limits of P1640. We refer to \cite{JustinSpeckle} for an example of a P1640 reduction when introducing the companions one at a time. Note  also that the exposure time can be lengthened and the other algorithm parameters can be finely tuned for better SNR. However we do not delve into the exploration of the whole parameter space: the results of such an optimization are presented in \citet{JustinSpeckle}. In this section we describe a couple of test cases which illustrate the different regimes of the regular and d-LOCI algorithms with regard to point source detectability.\\

\begin{figure}
\includegraphics[width=6in]{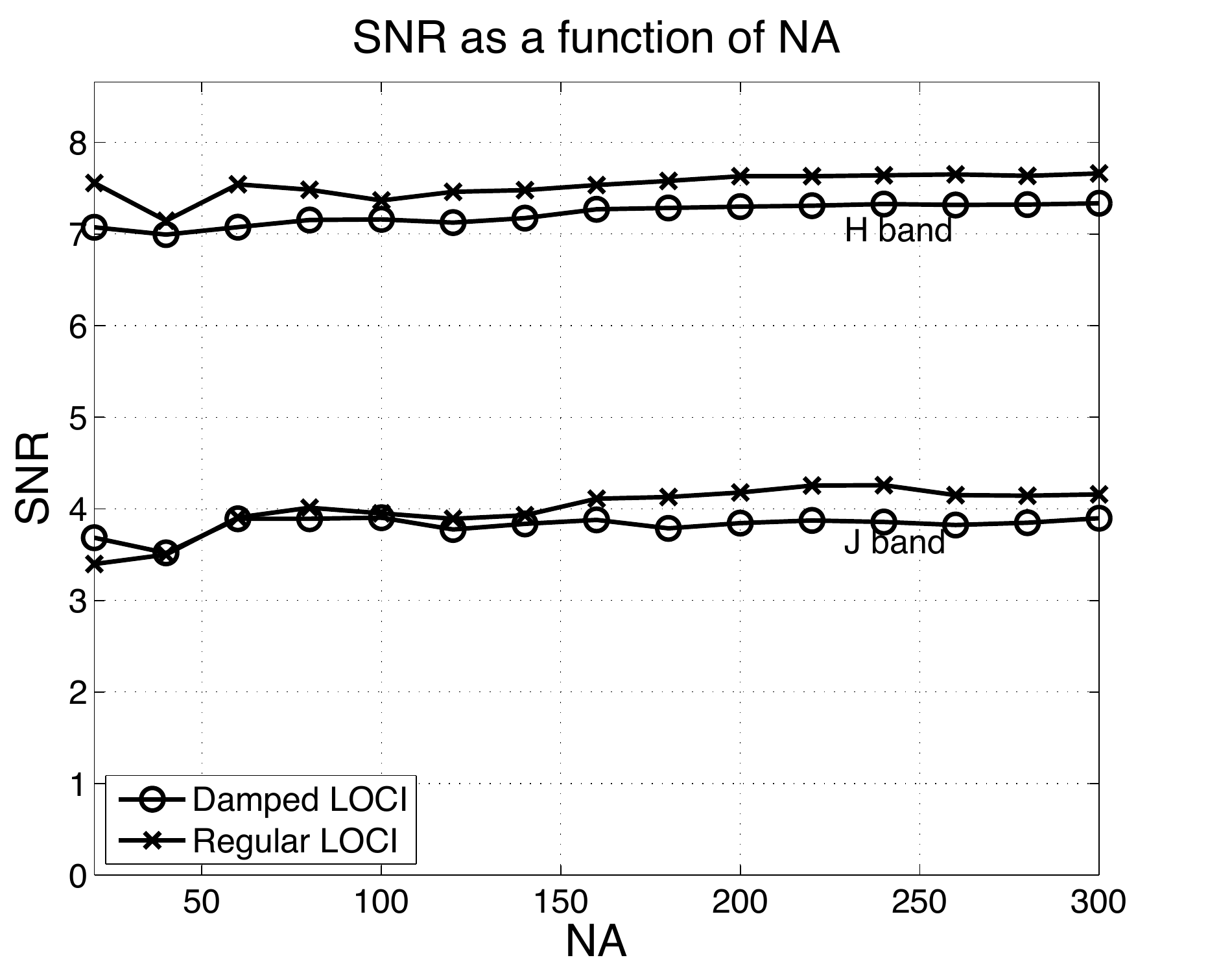}
\caption[SNR as a function of number of cubes]{Detectability of faint point sources in the case of an over-constrained LOCI reduction: SNR as a function of  the area of the optimization zone for constant exposure time and number of references (here the exposure time is $140$ s and $N_{\mathcal{R}} < 23$). The synthetic companions illustrated on Fig.~\ref{fig::SummaryFakePlanets} were used for these calculations: the figure shows the average of the post LOCI SNR over the 11 companions. Top curves: H-band results. This shows that in the over-constrained regime d-LOCI performs almost as efficiently as classical LOCI as far as the detectability of faint point sources is concerned. While these results were obtained using $g = 1$, reduction with $g \gg 1$ yield the same qualitative behavior, albeit at slightly higher SNR levels. Bottom curves: J-band results. Details about the second order trends in this figure are discussed in the body of the present manuscript.}
\label{fig::SNRbandOneFileNAChange20To300}
\end{figure}

We first consider the case of an over-constrained problem for which the number of free parameters, which scales with the number of reference frames, is smaller than the number of constraints, which scales with $N_{A}$, the number of PSF cores in the zone $\mathcal{O}$. We only reduce one data cube. This corresponds to a short integration, $140$ seconds, and a small number of references to choose from in order to build the calibration PSF: $N_{\mathcal{R}} < N_{\lambda} = 23$ (the actual value of $N_{\mathcal{R}}$ varies as a function of angular separation, at the location of the injected synthetic sources a typical value is $N_{\mathcal{R}}\sim 15 $). We fix the geometry of the problem as follows: $g = 1$, $dr = 2$ and $\delta = 1$. We then vary the size of the optimization zone by changing the value of $N_A$ from $20$ to $300$, and thus study the behavior of both LOCI and d-LOCI in the over-constrained case since $N_A > N_{\mathcal{R}}$ in these configurations. Fig.~\ref{fig::SNRbandOneFileNAChange20To300} illustrates how the average SNR of the $11$ injected companions, integrated over the wavelength channels in both J and H bands,  behaves in this configuration. The signal is estimated using aperture photometry at the known location of each artificial companion. The noise is estimated by computing the standard variation of the flux of a set of $100$ boxes surrounding the spiral of fake planets. Each photometric box has the area of a PSF core. Since the purpose of this section is to illustrate general trends pertaining to detectability with LOCI and d-LOCI, we  chose to only isolate two ``raw contrast'' levels: ``faint'' companions in  J band and ``bright'' companions in  H band, and only show how the averaged SNR over the 11 companions behaves of for these two sub-groups. A detailed study, with synthetic sources of continuously decreasing brightness, was carried out in \citet{JustinSpeckle} in order to derive the sensitivity limit of P1640. Both LOCI and d-LOCI present the same overall level of performance with slight upward SNR trend in both bands. In spite of having a penalty term that worsens the least-squares residual in the zone of the image where the PSF is subtracted, d-LOCI performs nearly as well than regular LOCI in the over-constrained case. The small SNR difference can be understood as follows: the noise suppression using pseudo-inverse methods is always more aggressive than when a damping term in introduced, and, in the over-constrained case, the likelihood of fitting companion flux with starlight is low. The upward trend in SNR as a function of the area of the optimization zone means, for the data-set studied here, that the higher companion throughput (e.g. larger signal) achieved by using a larger $N_A$ slightly dominates the degradation of the noise due the fact that the composite calibration image is calculated over a less local area.\\

\begin{figure}
\includegraphics[width=6in]{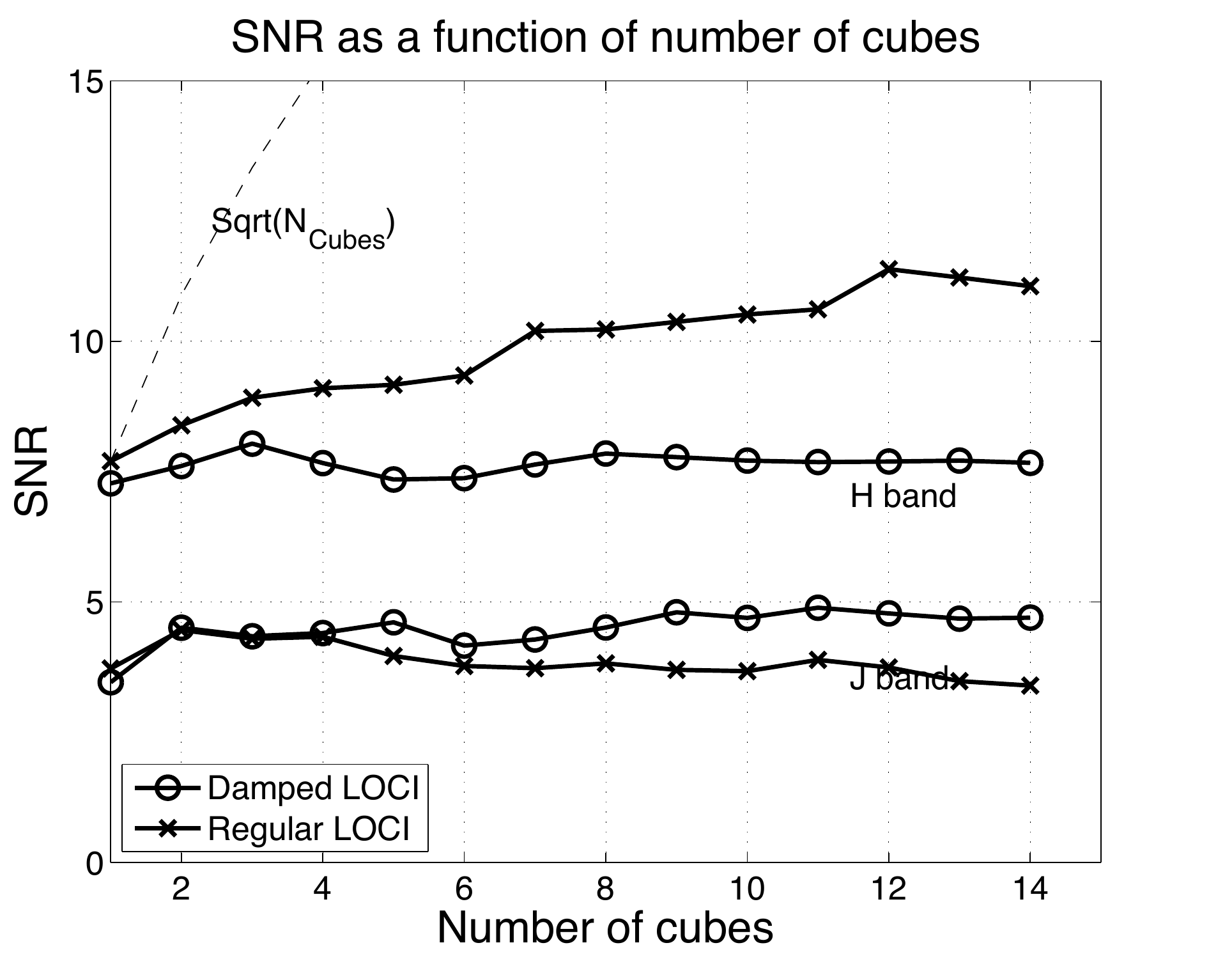}
\caption[SNR as a function of number of cubes]{Detectability of faint point sources in the case of an under-constrained LOCI reduction: SNR as a function of the number of cubes reduced (increasing exposure time and increasing number of references), for a constant area of the optimization zone, constant $N_A$. The synthetic companions illustrated on Fig.~\ref{fig::SummaryFakePlanets} were used for these calculations: the figure shows the average of the post LOCI SNR over the 11 companions.   In the case of under-constrained problems, the relative performances of classical LOCI and d-LOCI with respect to detectability  are a function of the contrast of the companion. Top curves: H-band results. For bright companions, H-band results, classical LOCI provides better noise suppression. Bottom curves: J-band results. For fainter companions, J-band, the noise suppression with classical LOCI is associated with a large signal depletion (e.g, the companion's signal is fitted by one of the references). When using d-LOCI, this depletion is mitigated, and thus  larger SNR are obtained. While these results were obtained using $g = 1$, reduction with $g \gg 1$ yield the same qualitative behavior, albeit at slightly higher SNR levels.}
\label{fig::SNRHbandNA300FilesChange1To15L}
\end{figure}

We then study the case of an under-constrained problem, for which the number of reference frames is larger than the number of PSF cores in the zone $\mathcal{O}$. To do so, we now fix the area of the optimization zone $N_A =100$ and vary the number of cubes reduced from $1$ to $14$. According to our frame selection strategy, illustrated on Fig.~\ref{Fig::TowerSortIllustration}, using more data-cubes should  increase the signal, since co-adding cubes increases the exposure time, and should also reduce the noise after LOCI, since $N_{\mathcal{R}}$ increases with the number of cubes, thus making the subtraction more efficient. This means that we expect the SNR to be an increasing function of $N_{\mathcal{R}}$. Our result are shown on Fig.~\ref{fig::SNRHbandNA300FilesChange1To15L}. Classical LOCI H-band results exhibit this behavior, with SNR increasing from $8$ to $11$, and yield better performances than d-LOCI reductions, for which the SNR remains constant at $8$. Note that the rise in detectability does not scale as $\sqrt{N_{\mathcal{R}}}$. This suggest that both our frame selection method (that was chosen for computational cost purposes)  our LOCI parameters are not fully optimized for the synthetic companions injected.  In the J-Band, for which the contrast is less favorable, d-LOCI leads to constant detectability levels ($SNR = 4.5$), while the performances of classical LOCI actually degrade with $N_{\mathcal{R}}$. In this configuration the likelihood of fitting companion flux is high, since there are more degrees of  freedom than there are constraints, and thus the algorithmic throughput decreases when $N_{\mathcal{R}}$ increases. When the companion is faint enough (in the J-band) then classical LOCI implementations that are under-constrained can reach a regime for which the companion's flux suppression is greater than the speckle noise attenuation.\\

The cases illustrated on Fig.~\ref{fig::SNRbandOneFileNAChange20To300} and Fig.~\ref{fig::SNRHbandNA300FilesChange1To15L} first confirm that when seeking for faint companions using LOCI reduction methods one should optimize the algorithm parameters beforehand using synthetic sources. In particular they show how the conditioning of the correlation matrix, $\mathbf{M}^{(R,R)}_{\mathcal{O}}$ is a critical metric in this parameter search and that $N_A$ and $N_{\mathcal{R}}$ ought to be adjusted so that the inverse problem is not under-constrained. This was carried out in the case of P1640 by \citet{JustinSpeckle}. Second, they show that d-LOCI detectability performances are close to the ones of classical LOCI. Overall, the damping makes d-LOCI less sensitive to the conditioning of the inverse problem, and his flat response to variations of the algorithm parameters might be used in order to discriminate astronomical companions from speckles residuals in the case of marginal detection. However a careful study of this second order trend is beyond the scope of the present paper. 
%


\section*{Acknowledgements}
We thank our anonymous referee for his/her very useful suggestions and Christian Marois for very productive conversations during the early stages of this manuscript. The research described in this publication was carried out in part at the Jet Propulsion Laboratory, California Institute of Technology, under a contract with the National Aeronautics and Space Administration. Project 1640 is funded by National Science Foundation grants AST-0520822, AST-0804417, and AST-0908484. This work was partially funded through the NASA ROSES Origins of Solar Systems Grant NMO710830/102190, the NSF AST-0908497 Grant. The adaptive optics program at Palomar is supported by NSF grants AST-0619922 and AST-1007046. LP was supported by an appointment to the NASA Postdoctoral Program at the JPL, Caltech, administered by Oak Ridge Associated Universities through a contract with NASA. LP and SH acknowledge support from the Carl Sagan Fellowship Program. 

\clearpage


\clearpage

\clearpage

\clearpage
\clearpage
\clearpage

\end{document}